\documentclass[10pt]{article}

\usepackage{amsmath}

\usepackage{array}

\usepackage{appendix}

\usepackage{tocloft}                   

\usepackage{graphicx}

\usepackage{amsfonts}

\usepackage{amssymb}

\usepackage{mathrsfs}

\usepackage{yfonts}

\usepackage{euscript}

\usepackage{centernot}                 

\usepackage{ifsym}                     

\usepackage{lmodern}                   

\usepackage{upgreek}

\usepackage{mathtools}

\usepackage{color}

\usepackage{slantsc}
\usepackage{calligra}

\usepackage{bbold}          

\usepackage[T1]{fontenc}

\usepackage{epsf}

\usepackage{latexsym}

\usepackage{tipa}

\usepackage{makeidx}

\makeindex

\def\b{\begin{equation}}

\def\e{\begin{equation}}

\def\be{\begin{equation}}              

\def\ee{\end{equation}}

\def\beq{\begin{equation}}

\def\eeq{\end{equation}}

\def\bea{\begin{eqnarray}}

\def\eea{\end{eqnarray}}

\def\half{\mbox{$\frac{1}{2}$}}

\def\m{\mbox{ }}

\def\mma {\m , \m \m }

\def\!{\hspace{-1.6667em}}

\def\n{\noindent}

\def\u{\underline}

\def\w{\widetilde}

\def\s{\stackrel}

\def\slTheta{\mathit{\Theta}}                     

\def\slLambda{\mathit{\Lambda}}                   

\def\slPhi{\mathit{\Phi}}                         

\def\uiR{\u{R}}

\def\uiq{\u{q}}

\def\uir{\u{r}} 

\def\bia{\mbox{\boldmath$a$}}

\def\bigg{\mbox{\boldmath$g$}}

\def\sbig{\mbox{\scriptsize\boldmath$g$}}

\def\bih{\mbox{\boldmath$h$}}

\def\bip{\mbox{\boldmath$p$}}

\def\biq{\mbox{\boldmath$q$}}

\def\bir{\mbox{\boldmath$r$}}

\def\biM{\mbox{\boldmath$M$}}

\def\biO{\mbox{\boldmath$O$}}

\def\biP{\mbox{\boldmath$P$}}

\def\biQ{\mbox{\boldmath$Q$}}

\def\sbim{\mbox{\scriptsize\boldmath$m$}}

\def\sbin{\mbox{\scriptsize\boldmath$n$}}

\def\sbiM{\mbox{\scriptsize\boldmath$M$}}

\def\sbiN{\mbox{\scriptsize\boldmath$N$}}

\def\sbiQ{\mbox{\scriptsize\boldmath$Q$}}

\def\birho{\mbox{\boldmath$\rho$}}

\def\brho{\birho}                                   

\def\mB{\mbox{B}}  

\def\mC{\mbox{C}}                        

\def\mD{\mbox{D}}                        

\def\mE{\mbox{E}}                        

\def\mI{\mbox{I}}                        

\def\mM{\mbox{M}}                        

\def\mN{\mbox{N}}

\def\mR{\mbox{R}}                        

\def\mS{\mbox{S}}                        

\def\mT{\mbox{T}} 

\def\mb{\mbox{b}}

\def\me{\mbox{e}}

\def\mg{\mbox{g}}

\def\mh{\mbox{h}}

\def\mp{\mbox{p}}

\def\ms{\mbox{s}}

\def\mt{\mbox{t}}

\def\bh{\u{\u{\mbox{h}}}  }            

\def\suupbeta{\mbox{\scriptsize$\u{\upbeta}$}}

\def\bM{\mbox{\bf M}}

\def\bN{\mbox{\bf N}}

\def\bg{\mbox{\bf g}}

\def\bh{\mbox{\bf h}}

\def\bm{\mbox{\bf m}}

\def\bn{\mbox{\bf n}}

\def\bp{\mbox{\bf p}}

\def\bupSigma{\mbox{\boldmath$\Sigma$}}                 
\def\sbupSigma{\mbox{\scriptsize\boldmath$\Sigma$}}     
\def\sbSigma{\mbox{\scriptsize\boldmath$\Sigma$}}

\def\bcalD{\mbox{\boldmath ${\cal D}$}}

\def\bcalK{\mbox{\boldmath ${\cal K}$}}

\def\fG{\mbox{\sffamily G}}

\def\fL{\mbox{\sffamily L}}

\def\sa{\mbox{\scriptsize a}}

\def\sb{\mbox{\scriptsize b}}

\def\scc{\mbox{\scriptsize c}}

\def\sd{\mbox{\scriptsize d}}

\def\se{\mbox{\scriptsize e}}

\def\sg{\mbox{\scriptsize g}}

\def\si{\mbox{\scriptsize i}}

\def\sm{\mbox{\scriptsize m}}

\def\sn{\mbox{\scriptsize n}} 

\def\so{\mbox{\scriptsize o}}

\def\sr{\mbox{\scriptsize r}}

\def\sss{\mbox{\scriptsize s}}  

\def\st{\mbox{\scriptsize t}}

\def\sv{\mbox{\scriptsize v}}

\def\sA{\mbox{\scriptsize A}} 

\def\sB{\mbox{\scriptsize B}}

\def\sC{\mbox{\scriptsize C}}

\def\sD{\mbox{\scriptsize D}}

\def\sE{\mbox{\scriptsize E}}

\def\sG{\mbox{\scriptsize G}}

\def\sH{\mbox{\scriptsize H}}

\def\sJ{\mbox{\scriptsize J}}

\def\sL{\mbox{\scriptsize L}} 

\def\sM{\mbox{\scriptsize M}}

\def\sP{\mbox{\scriptsize P}}

\def\sR{\mbox{\scriptsize R}}

\def\sS{\mbox{\scriptsize S}}

\def\sW{\mbox{\scriptsize W}}

\def\sfA{\mbox{\sffamily{\scriptsize A}}}     

\def\sfB{\mbox{\sffamily{\scriptsize B}}}     

\def\sbM{\mbox{{\bf \scriptsize M}}}

\def\sbN{\mbox{{\bf \scriptsize N}}}

\def\sbcP{\mbox{\boldmath \scriptsize ${\cal P}$}}

\def\sbcF{\mbox{\boldmath \scriptsize ${\cal F}$}}

\def\sbcG{\mbox{\boldmath \scriptsize ${\cal G}$}}

\def\sbcL{\mbox{\boldmath \scriptsize ${\cal L}$}}

\def\sbcS{\mbox{\boldmath \scriptsize ${\cal S}$}}

\def\sbcM{\mbox{\boldmath \scriptsize ${\cal M}$}}

\def\barp{\bar{\tt p}}

\def\cr{\mbox{\scriptsize{\bf $\m  \times \m $}}}

\def\sumi2{\sum\mbox{}_{\mbox{}_{\mbox{\scriptsize $i$=1}}}^2}

\def\sumi3{\sum\mbox{}_{\mbox{}_{\mbox{\scriptsize $i$=1}}}^3}

\def\sumABcycles3{\sum\mbox{}_{\mbox{}_{\mbox{\scriptsize cycles $A,B$=1}}}^{3}}

\def\sumCDcycles3{\sum\mbox{}_{\mbox{}_{\mbox{\scriptsize cycles $C,D$=1}}}^{3}}

\def\sumdn{\sum\mbox{}_{\mbox{}_{\mbox{\scriptsize $I$=1}}}^{n \, d}}

\def\sumIN{\sum\mbox{}_{\mbox{}_{\mbox{\scriptsize $I$=1}}}^{N}}

\def\sumj3{\sum\mbox{}_{\mbox{}_{\mbox{\scriptsize $j$=1}}}^3}

\def\sumk3{\sum\mbox{}_{\mbox{}_{\mbox{\scriptsize $k$=1}}}^3}

\def\prodiA1{\prod\mbox{}_{\mbox{}_{\mbox{\scriptsize $i$=1}}}^{A - 1}}

\def\d{\textrm{d}}                                                  

\def\pa{\partial}                                                   

\def\Circ{\mbox{\Large$\circ$}}                                     

\def\es{\m = \m}

\def\:={\m := \m}

\def\=:{\m =: \m}

\def\Abs{\mbox{\Large $\mathfrak{a}$}\mb\ms}                         

\def\FrS{\mbox{\Large $\mathfrak{s}$}}                         
\def\lFrm{\mbox{\large$\mathfrak{m}$}}                         
\def\FrM{\mbox{$\mathfrak{M}$}}                                
\def\lFrg{\mbox{\Large$\mathfrak{g}$}}                         
\def\nFrg{\mbox{\large$\mathfrak{g}$}}                         

\def\Hilb{\mbox{{\boldmath$\mathfrak{H}$}ilb}}                 

\def\scC{\mbox{\scriptsize ${\cal C}$}}                    

\def\scD{\mbox{\scriptsize ${\cal D}$}}                    

\def\scE{\mbox{\scriptsize ${\cal E}$}}                    

\def\scH{\mbox{\scriptsize ${\cal H}$}}                    

\def\scM{\mbox{\scriptsize ${\cal M}$}}                    

\def\scQ{\mbox{\scriptsize ${\cal Q}$}}                    

\def\bLin{\sbcL\mbox{\bf in}} 

\def\bFlin{\sbcF\mbox{\bf lin}} 

\def\Quad{\scQ\mbox{uad}}                                  

\def\Chronos{\scC\mbox{hronos}}                            

\def\bGauge{\sbcG\mbox{\bf auge}} 

\def\bShuffle{\sbcS\mbox{\bf huffle}} 

\def\FrQ{\mbox{\Large $\mathfrak{q}$}}                               
\def\bFrC{\mbox{\boldmath$\mathfrak{C}$}}                            
\def\Phase{\mbox{{\boldmath$\mathfrak{P}$}hase}}                     

\def\bFrR{\mbox{\boldmath$\mathfrak{R}$}}                            
\def\Rig-Phase{\bFrR\mbox{ig-}\Phase}                                
                                                                													   
\def\lFrr{\mbox{\Large $\mathfrak{r}$}}                              

\def\FrP{\mbox{\Large $\mathfrak{p}$}}                                 
\def\FrR{\mbox{\boldmath$\mathfrak{R}$}}                             

\def\bFrR{\mbox{\boldmath$\mathfrak{R}$}}                            

\def\bFrR{\mbox{\boldmath$\mathfrak{R}$}}                            

\def\1mat{\u{\u{1}}}                                                 

\def\Positive-Modespace{\mbox{{\boldmath$\mathfrak{M}$}odespace$^+$}}

\def\POSITIVE-MODESPACE{\mbox{{\boldmath$\mathfrak{M}$}ODESPACE$^+$}}
                                                                                                                             														
\def\bFrS{\mbox{\Large $\mathfrak{s}$}}                              
			
\def\Riem{\bFrR\mbox{iem}}                                           
\def\CRiem{\bFrC\Riem}                                               

\def\Superspace{\bFrS\mbox{uperspace}}                               
\def\CS{\bFrC\bFrS}                                                  

\def\lE{\mbox{\Large E}} 

\def\lS{\mbox{\Large S}}

\def\Kin-Hilb{\mbox{{\boldmath$\mathfrak{K}$}in-\Hilb}}                     

\def\Mid-Hilb{\mbox{{\boldmath$\mathfrak{M}$}id-\Hilb}}                     

\def\Dyn-Hilb{\mbox{{\boldmath$\mathfrak{D}$}yn-\Hilb}}                     

\def\5Star{\mbox{\Large$\star$}}              

\def\K{Kucha\v{r} }

\def\Frr{\mbox{$\mathfrak{r}$}}

\begin{document}

\begin{center}

\Huge{\bf A LOCAL RESOLUTION OF}

\vspace{.1in}

\normalsize

\Huge{\bf THE PROBLEM OF TIME}

\vspace{.15in}

\large{\bf II. Configurational Relationalism via a generalization of Group Averaging} 

\vspace{.15in}

{\large \bf E.  Anderson}$^1$ 

\vspace{.15in}

{\large \it based on calculations done at Peterhouse, Cambridge} 

\end{center}

\begin{abstract}

In this article, we consider a second Problem of Time facet. 
This started life as Wheeler's Thin Sandwich Problem, within the narrow context of  
I) GR-as-Geometrodynamics, in particular its momentum constraint.   
II) A Lagrangian variables level treatment.    
Conceiving in terms of Barbour's Best Matching is a freeing from I), now in the context of first-class linear constraints. 
Conceiving in terms of the underlying Background Independence aspect -- the titular Configurational Relationalism -- serves moreover to remove II) as well. 
This is implemented by the $\lFrg$-act, $\lFrg$-all method. 
I.e.\ given an object $O$ that is not $\lFrg$-invariant, we act on it with $\lFrg$ and then apply an operation involving the whole group.
This has the effect of double-cancelling the introduction of our group action, thus yielding a $\lFrg$-independent version of $O$.
A first example of whole-group operation is group averaging. 
This is a valuable prototype by its familiarity to a large proportion of Mathematics and Physics majors 
through featuring in elementary Group Theory and Representation Theory courses (and which can be traced back to Cauchy). 
In particular, this is much more familiar than Thin Sandwiches or Best Matching!   
Secondly, extremization over the group, of which Best Matching, and taking infs or sups over the group, are examples. 
Some further significant examples of this method in modern geometry and topology include those of Kendall, Younes, Hausdorff and Gromov.     
In this way, we populate this approach with examples well beyond the usual GR literature's by DeWitt, Barbour and Fischer (which we also outline).

\end{abstract}

$^1$ dr.e.anderson.maths.physics *at* protonmail.com

\section{Introduction}\label{Intro}

In this article, we consider a second Problem of Time facet. 

\m 

\n This started life as Wheeler's {\bf Thin Sandwich Problem} \cite{BSW, WheelerGRT}, within the following narrow context.

\m 

\n{\bf Restriction 1)} GR-as-Geometrodynamics (\cite{ADM, DeWitt67}, reviewed in e.g.\ \cite{MTW, Gour, Giu15} and Chapter 8 of \cite{ABook} and for which Sec 2 provides an outline).  

\m 

\n{\bf Restriction 2)} Thin Sandwich Problem (Sec 3) therein concerns the momentum constraint, to be treated moreover at the level of Lagrangian variables.  

\m 

\n The Thin Sandwich Problem was then listed as one of the Problem of Time 
\cite{Battelle, DeWitt67, Dirac, K81, K91, K99, APoT, FileR, APoT2, AObs, APoT3, ALett, ABook,  A-CBI, I, III, IV} facets  
in Kucha\v{r} and Isham's reviews \cite{K92, I93}.  

\m 

\n It is however better for this Problem of Time facet, and underlying Background Independence aspect, to be set free from restrictions 1) and 2).  

\m 

\n{\bf Liberation 1)} {\bf The Best Matching Problem} reconceptualization (Sec 4) frees us from restriction to GR-as-Geometrodynamics.
While this was first proposed -- by Barbour and Bertotti -- \cite{BB82} a decade prior to \K and Isham's reviews, 
almost all work with it postcedes these reviews \cite{B94I, RWR, B03, ABFO, FileR, APoT2, APoT3, ABook}.  
Best Matching involves group-correcting Principles of Dynamics actions with respect to $\lFrg$ actions and then extremizing over $\lFrg$. 
Whatever first-class linear constraints thereby arise thus are to be treated at the level of Lagrangian variables 
as equations from which to eliminate the $\lFrg$-auxiliary variables that implement the group corrections. 

\m 

\n In the case of GR-as-Geometrodynamics, the group 

\n\be 
\lFrg = Diff(\bupSigma)  \m :
\ee 
the spatial diffeomorphisms on a fixed spatial topological manifold $\bupSigma$.  
The ensuing first-class linear constraint is indeed the GR momentum constraint.
The above Thin Sandwich Problem is thus indeed recovered in this arena.   

\m 

\n{\bf Liberation 2)} Freedom from being a Lagrangian-level (and thus a fortiori classical-level) venture. 
This is attained by Configurational Relationalism \cite{FileR, APoT2, APoT3, ABook} (Sec 5). 
E.g.\ this can also apply at the Hamiltonian level, 
                          to conformally-transformed constraints, 
					   or at the level of solving the quantum equations. 
It can also apply moreover to whatever other objects one's theory needs: notions of distance, information, correlation, quantum operators...

\m  					   

\n{\bf The Direct Method} (Sec 6), on the one hand, 
is an occasionally available implementation, in close analogy with using gauge-invariant variables in those fortunate few examples possessing such. 
					   
\m 					   
					   
\n {\bf The Indirect} $\lFrg${\bf -act} $\lFrg${\bf -All Method} (Sec 7), on the other hand, 
is a universal implementation of Configurational Relationalism.   
Here incipient $\lFrg$-noninvariant objects are firstly acted upon by $\lFrg$, 
followed by applying an operation involving the whole of $\lFrg$ is performed to produce a $\lFrg$-invariant version of the object. 

\m 

\n Best Matching can now be viewed as a subcase of the latter with extremization over $\lFrg$ as its $\lFrg$-all operation.

\m 

\n Once one has this conception, moreover, 'group averaging' makes for a far more useful prototype. 
For, out of featuring in elementary undergraduate Group Theory and Representation Theory, it is much more widely known than Best Matching.
Group averaging can in fact be traced back to Cauchy \cite{Cauchy}, 
though widespread knowledge of it had to await Burnside's rather later exposition \cite{Burnside}.  

\m 

\n Taking infs or sups over $\lFrg$ is a further example of note of extremizing $\lFrg$-all operation.  
Sec 8 subsequently provides a further suite of high-profile examples of $\lFrg$-act $\lFrg$-all methods. 
These are familiar in Theoretical Statistics \cite{Kendall}, Geometry \cite{Younes} and Topology \cite{Gromov}, 
as well as some occurrences in earlier GR and Problem of Time literature \cite{DeWitt67, Fischer70}.  

\m 

\n All in all, Configurational Relationalism thus in fact has well-known, and elsewhere widely useful foundations. 
It is, in particular, readily accessible to Mathematics or Theoretical Physics majors via the group averaging prototype that the current Article emphasizes.

\section{Dynamical formulation of GR}\label{Dyn-GR}

\subsection{Spacetime action for GR}

{\bf Structure 1} The action for GR in its most commonly encountered spacetime formulation is the {\it Einstein--Hilbert action}\footnote{This Series uses 
$8 \, \pi \, G = 1 = c$ `fundamental units'. }
\beq
{\cal S}^{\sG\sR}_{\sE\sH}  \es  \int_{\lFrm} \d^4 X \sqrt{|\mg|} \, {\cal R}^{(4)}(\vec{X}; \bg]  \m .  
\label{S-EH}  
\eeq
$\vec{X}$ with components $X^{\mu}$          are here spacetime coordinates, and 
$\bg$ with components $\mg_{\mu\nu}$ is the spacetime metric, with determinant $\mg$ and spacetime Ricci scalar ${\cal R}^{(4)}$.  
To include the cosmological constant $\slLambda$, ${\cal R}^{(4)}$ is corrected by $- 2 \, \slLambda$. 

\m 

\n{\bf Aside 1} GR has more Background Independence aspects than theories of Mechanics do \cite{APoT3}.
This follows from GR possessing a nontrivial notion of spacetime, which geometrizes a wider range of features than Mechanics' notion of split space-time does.
The latter is far more of a composition of separate notions of space and time: multiple copies of a spatial geometry strung together by a time direction, 
whereas the former is a co-geometrization of space and time. 
See Articles III, X and XII for these further aspects.
A space-time split can nonetheless be applied to GR, as follows.

\subsection{The ADM split}

\n{\bf Structure 1} Arnowitt--Deser--Misner (ADM) \cite{ADM} split the spacetime metric $\mg_{\mu\nu}$ into induced metric $\bh$ with components $\mh_{ij}$, 
shift $\u{\upbeta}$ and lapse $\upalpha$ pieces: 
\beq
\mg_{\mu\nu}  \es
\left(
\stackrel{    \mbox{$ \upbeta_{k}\upbeta^{k} - \upalpha^2$}    }{ \m  \m  \upbeta_{j}    } \stackrel{    \mbox{$\upbeta_{i}$}    }{  \m \m  \mh_{ij}    }
\right)              \m . 
\label{ADM-split}
\eeq
See Fig \ref{ADM-Split} in this regard. 
This split is often presented for a foliation, though two infinitesimally close hypersurfaces suffices.   
%

\m 

\n{\bf Modelling Assumption} Space $\bupSigma$ is here a hypersurface -- codimension-1 -- and positive-definite at the level of metric geometry.
We take space to be compact without boundary (CWB) for simplicity, and also connected.  
A fixed $\bupSigma$ is to be shared by {\sl all} the spatial configurations in a given Geometrodynamics.  
I.e.\ dynamical formulations of GR such as Geometrodynamics are built subject to the restriction of not allowing for topology change. 
We moreover take $\bupSigma = \mathbb{S}^3$ when we need to be specific.

{            \begin{figure}[!ht]
\centering
\includegraphics[width=0.55\textwidth]{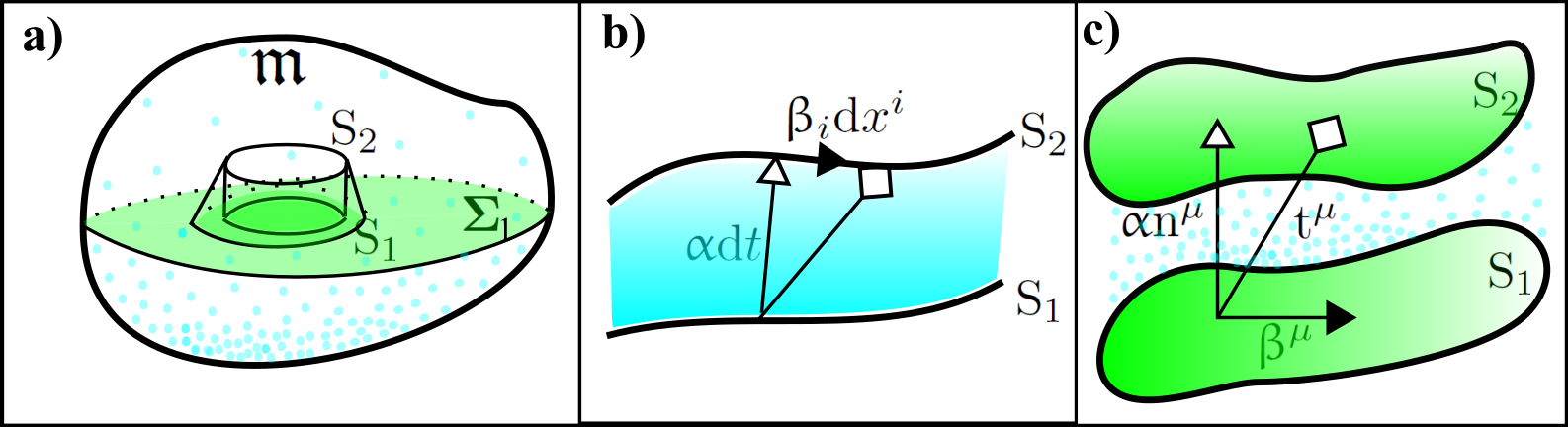}
\caption[Text der im Bilderverzeichnis auftaucht]{        \footnotesize{Arnowitt--Deser--Misner (ADM) 3 + 1 split of a region of spacetime, 
with lapse $\upalpha$ and shift $\u{\upbeta}$.
In the current Series, green manifolds denote space and blue ones denote spacetime. 
    White triangular   arrows indicate time elapsed, 
    black triangular   arrows indicate diffeomorphism shifts within a given spatial slice, 
and white spear-headed arrows indicate point-identification maps between adjacent spatial slices.  
$\sS_1$ and $\sS_2$ are pieces of two spatial hypersurfaces $\bupSigma_1$ and $\bupSigma_2$.}  }
\label{ADM-Split}\end{figure}          }

\m 

\n{\bf Structure 2} The corresponding split of the inverse metric is  
\beq
\mg_{\mu\nu}  \es 
\left(
\stackrel{    \mbox{$ - {1}/{\upalpha^2}$}    }{ \m  \m   {\upbeta^j}/{\upalpha^2}    }
\stackrel{    \mbox{$   {\upbeta^i}/{\upalpha^2}$}    }{  \m \m  \mh^{ij}  - {\upbeta^i\upbeta^j}/{\upalpha^2}  }
\right)                                                                                                             \m .
\eeq
That of the square-root of the determinant is
\be 
\sqrt{|\mg|} = \upalpha\sqrt{\mh}                                                                                   \m .
\ee

\subsection{ADM action}

{\bf Proposition 1} The Einstein--Hilbert action correspondingly splits into the ADM action 
\beq
{\cal S}_{\sG\sR}^{\sA\sD\sM}  \m \propto \m  \int \d t \int_{\sbupSigma}\d^3 x \, {\cal L}_{\sG\sR}^{\sA\sD\sM}  
                               \es       \int \d t \int_{\sbupSigma}\d^3 x \sqrt{\mh} \, \upalpha \{ {\cal K}_{ab} {\cal K}^{ab} - {\cal K}^2   +  {\cal R} - 2 \, \slLambda \}  \m ,
\label{S-ADM}  
\eeq
where $x^i$ are spatial coordinates, 
$t$        is coordinate time, 
$\mh$ is $\bh$'s determinant and ${\cal R}$ its Ricci scalar, and 
$\bcalK$   is the extrinsic curvature of space within ambient spacetime.

\m 

\n{\u{Outline of Proof}} Use a combination of contractions of the Gauss and Ricci equations \cite{Wald} of extrinsic geometry, 
alongside discarding a total divergence since $\bupSigma$ is CWB.  $\Box$

\subsection{ADM-Lagrangian action}\label{DeWitt-supermetric}

{\bf Proposition 2} In terms of GR-as-Geometrodynamics' Lagrange-type configuration--velocity variables
\be
(  \bh, \m \bh^{\prime})                                                                                       \m ,  
\ee
the ADM action can be recast as 
\be 
{\cal S}_{\sG\sR}^{\sA\sD\sM\mbox{$-$}\sL}  \propto  \int \d \mt \int_{\sbSigma} \d^3 x \, {\cal L}_{\sG\sR}^{\sA\sD\sM\mbox{$-$}\sL}  
                                              \es    \int \d \mt \int_{\sbSigma} \d^3 x \sqrt{\mh} \, \upalpha 
\left\{
\frac{{\cal T}_{\sG\sR}^{\sA\sD\sM\mbox{$-$}\sL}}{4 \, \upalpha^2} +  {\cal R} - 2 \, \slLambda
\right\}                                                                                                       \m .    
\ee
The {\it GR kinetic term} here is  
\beq
{\cal T}_{\sG\sR}^{\sA\sD\sM\mbox{$-$}\sL}  \:=  {||\delta_{\suupbeta}\bh||_{\sbM}}^2 
                                            \es  {||\{\mbox{}^{\prime} - \pounds_{\u{\upbeta}}\} \bh ||_{\sbM}}^2
                                            \es  \mM^{abcd} 
\left\{ 
\mh^{\prime}_{ab}  -  \pounds_{\u{\upbeta}} \mh_{ab} 
\right\}
\left\{ 
\mh^{\prime}_{cd}  -  \pounds_{\u{\upbeta}} \mh_{cd}  
\right\}                                                                                                 \m ,  
\ee 
where 
\be 
\mM^{abcd}  \:=  \sqrt{\mh} \{ \mh^{ac} \mh^{bd} - \mh^{ab} \mh^{cd} \}                                  \m . 
\ee 
This is established via the following Lemma. 

\m 

\n{\bf Lemma 1} (ADM extrinsic curvature--metric velocity relation) 
\beq
\mbox{\boldmath ${\cal K}$}    \es  \frac{  \updelta_{\vec{\upbeta}}{\bh}  }{  2 \, \upalpha  }  
                               \es  \frac{  \{\mbox{}^{\prime} - \pounds_{\u{\upbeta}}\} \bh  }{  2 \, \upalpha  }      \m ,   
\label{Kij-ADM}
\eeq
or, in terms of components, with a useful further expression,
\beq
{\cal K}_{ij}  \es  \frac{\updelta_{\vec{\upbeta}}{\mh}_{ij}}{2 \, \upalpha}
               \es  \frac{\mh_{ij}^{\prime} - \pounds_{\underline{\upbeta}}\mh_{ij}}{2 \, \upalpha} 
               \es  \frac{\mh_{ij}^{\prime} - 2 \, D_{(i}\upbeta_{j)}}{2 \, \upalpha}                          \m .   
\label{Kij-ADM2}
\eeq			   
$\updelta_{\vec{\upbeta}}$ is here {\it Kucha\v{r}'s hypersurface derivative}, $D_i$                      the          spatial covariant derivative, and  
\be 
\mbox{}^{\prime}  \:=  \frac{\pa}{\pa \mt}                                                              \m , 
\ee 
\n{\bf Remark 1} $\bM$ is here conceptually the {\it GR configuration space metric}, which arose historically as the {\it inverse DeWitt supermetric} \cite{DeWitt67}. 

\m 

\n{\bf Remark 2} GR thus places a metric on the configuration space $\Riem(\bupSigma)$; see Article VI for more about the structure of this space.

\m 

\n{\bf Remark 3} {\sl Supermetric} refers to its possessing four indices and already being built out of one preceding notion of metric, $\mh_{ab}$. 
Ab initio, this is a 4-index inner product. 
DeWitt's \cite{DeWitt67} 2-index to 1-index map 
\be 
\mh_{ab} \mapsto \mh^A
\ee 
moreover recasts it in the standard form for a metric, with two downstairs indices:  
\be 
\mM^{abcd} \mapsto \mM_{AB}  \m .
\ee  
[In this context, the capital Latin indices run from 1 to 6.]

\m 

\n{\bf Remark 4} Pointwise, this is a $(-+++++)$ metric. 
It is thus an infinite-dimensional analogue of semi-Riemannian metric.   
This `DeWittian' indefiniteness is associated with the expansion of the Universe giving a negative contribution to the GR kinetic energy.  
This is unrelated to the much better-known Lorentzian indefiniteness of SR and GR spacetimes themselves.

\m

\n{\bf Corollary 1} The manifest `(configuration-space) geometrical DeWitt action' is  
\be 
{\cal S}_{\sG\sR}^{\sD\se\sW\si\st\st}  \es  
\int \d \mt \int_{\sbSigma} \d^3 x \sqrt{\mh}\,\upalpha 
\left\{
\frac{{||\delta_{\suupbeta}\bh||_{\mbox{\scriptsize ${\bM}$}}}^2  }{4 \, \upalpha^2}  +  {\cal R} - 2 \, \slLambda
\right\}                                                                                                             \m . 
\ee

\subsection{GR's momenta}

{\bf Proposition 3} The GR-as-Geometrodynamics moment bear the following close relation to the extrinsic curvature.  
\beq
\mp^{ij}  \:=  \frac{\updelta {\cal L}_{\sG\sR}^{\sA\sD\sM\mbox{$-$}\sL}}{\updelta \dot{\mh}_{ij}}  
          \es  \sqrt{\mh}\{ {\cal K}^{ij} - {\cal K} \, \mh^{ij} \} 
		  \:=  \mM^{ijkl} \frac{\updelta_{\vec{\upbeta}}\mh_{kl}}{2 \, \upalpha}                     \m .  
\label{Gdyn-momenta-2}
\eeq
Or, in coordinate-free notation, 
\beq
\u{\bp}       \:=  \frac{\updelta {\cal L}_{\sG\sR}^{\sA\sD\sM\mbox{$-$}\sL}}{\updelta \dot{\u{\bh}}}  
          \es  \sqrt{\mh}\{ \u{\bcalK} - {\cal K} \, \u{\bh} \} 
		  \:=  \u{\u{\bM}} \cdot \frac{\updelta_{\vec{\upbeta}}\u{\bh}}{2 \, \upalpha}                           \m .  
\label{Gdyn-momenta}
\eeq
{\bf Remark 1} GR's momenta can thus be described as densitized ${\cal K}_{ij}$ with a particular trace term subtracted off.  

\m 

\n{\bf Corollary 2}
\beq
\mp = - 2 \sqrt{\mh} \, {\cal K}                                                   \m . 
\label{Tr}
\eeq
%

\subsection{GR's constraints}\label{GR-cons-and-sp-diff}

{\bf Proposition 4} The ADM-Lagrangian action encodes the following constraints. 
\beq
\mbox{\it GR Hamiltonian constraint } \m  \scH  \:=  {||\mp||_{\sbN}}^2  - \sqrt{\mh}\{  {\cal R} - 2 \, \slLambda\}   
                                                \es  \mN_{ijkl}\mp^{ij}\mp^{kl} - \sqrt{\mh}\{  {\cal R} - 2 \, \slLambda\} 
                                                \es  0                                                                        \m , 
\label{Hamm}
\eeq 
for 
\be
\mN^{AB}  \es  \mN_{abcd}  
          \es  \frac{1}{\sqrt{\mh}} \{  \mh_{ac}\mh_{bd} - \half \mh_{ab}\mh_{cd}  \}                           \m , 
\ee
\beq
\mbox{\it GR momentum constraint } \m \u{\sbcM} \:= -2 \u{\bcalD} \, \u{\u{\bp}} = 0 \m \mbox{ with components } \m 
                                      \scM_{i}  \es  - 2 \, {\cal D}_{j} {\mp^{j}}_{i} 
                                                \es                   0                                                       \m .  
\label{Momm}
\eeq
{\u{Proof}} These follow from variation with respect to the lapse $\upalpha$ and shift $\u{\upbeta}$ respectively.  $\Box$ 
 
\m  
 
\n{\bf Remark 1} A conceptual name for this $\bN$ the {\it GR constraint metric}, whereas a historical one is the {\it DeWitt supermetric} itself \cite{DeWitt67}.  
(This reflects constraint rather than action primality in the Wheeler school \cite{Battelle}.)  

\m 

\n{\bf Remark 2}  The GR momentum constraint is linear in the momenta.
It can be straightforwardly interpreted as physicality residing not in the 3 degrees of freedom per space point choice of point-identification 
but rather solely in terms of the 3-metric's {\sl other} $6 - 3 = 3$ degrees of freedom.
This amounts to the diffeomorphism-invariant information in the 3-metric,   
corresponding to the {\it 3-geometry} (named for the 3-dimensionality of space, not for still having 3 degrees of freedom per space point). 

\m 

\n{\bf Structure 1} GR is thereby less redundantly a dynamics of 3-geometries \cite{Battelle, DeWitt67} on the quotient configuration space,  Wheeler's \cite{Battelle}
\beq
\Superspace(\bupSigma)  \:=  \frac{\Riem(\bupSigma)}{Diff(\bupSigma)}                                                          \m . 
\label{Superspace}
\eeq
See Article V.I for further details and \cite{DeWitt67, DeWitt70, Fischer70, FM96, Giu09, Giu15} for original technical references.  
 
\m 

\n{\bf Remark 3} The GR Hamiltonian constraint is more involved to interpret.  
It is `purely-quadratic in the momenta', meaning it consists of a quadratic form plus a zero-order piece but with no linear piece:
\beq
\Quad  \:=  \half {|| \biP||_{\sbiN}}^2  - W(\biQ)  
       \es  \half N^{\sfA\sfB}(\biQ) P_{\sfA} P_{\sfB} - W(\biQ)  
       \es  0                                                     \m .  
\label{Quad}
\eeq
We shall see in Sec IV.3 that this property leads to the Frozen Formalism Facet of the Problem of Time; 
this property of course holds for $\Chronos$ more generally.   

\m 

\n{\bf Remark 4} Wheeler asked the following question, which readily translates to asking for zeroth principles `constraint provider' reasons for the form of the 
crucial GR Hamiltonian constraint, $\scH$.
\beq
\stackrel{\stackrel{\mbox{\it \normalsize ``If one did not know the Einstein--Hamilton--Jacobi equation,}} 
         {\mbox{\it \normalsize             how might one hope to derive it straight off from plausible first principles}}}
         {\mbox{\it \normalsize             without ever going through the formulation of the Einstein field equations themselves?"}}		 
\label{Wheeler-Q}
\eeq
This is in the context of no longer considering just the Geometrodynamics specific to GR, but rather a multiplicity of geometrodynamical theories,   
and is furthermore an appeal to seek for a {\sl selection principle} that picks out the GR case. 
We shall name answers to this question in Sec III.4.2, and detail them in Articles IX and XII.  

\m 

\n{\bf Remark 5} GR's phase space degrees of freedom count works out as 
$ 10 \times 2$ (including the lapse $\upalpha$ and shift $\u{\upbeta}$ as well as $\bh$) 
$- 3 \times 2$ (due to the shift being a Lagrange multiplier, so its momentum is zero) 
$- 1 \times 2$ (by     the lapse being a Lagrange multiplier, so its momentum is zero also) 
$- 3 \times 2 - 1 \times 2 = 2 \times 2$.

\subsection{Grounding Article I's minisuperspace metrics}

In minisuperspace, full GR's $\bM(\bh(\underline{x}))$ has collapsed to an ordinary $6 \times 6$ matrix, $\biM(\bih)$. 
This is an overall -- rather than independently per space point -- curved ($-+++++$) `minisupermetric'.  
Two simpler subcases nested within this are as follows.  

\m 

\n i) Diagonal Minisuperspace involves a yet smaller $3 \times 3$ ($-++$) matrix $\biM(\bih)$ \cite{Magic}.  

\m 

\n ii) Isotropic Minisuperspace: flat single-number (--) minisupermetric, for instance for $\bupSigma = \mathbb{S}^3$ with standard hyperspherical metric.  
This is a closed cosmological model, and simpler than i) through not modelling anisotropy.

\section{The Thin Sandwich Problem (precursor of facet 0b)}\label{TSP}

{\bf Remark 1} Solve the Lagrangian form of the constraints for $\upalpha$ and $\upbeta$. 
$\scH$ is algebraic in $\upalpha$, giving
\be 
\upalpha \es  \frac{1}{2} \sqrt{ \frac{\mT_{\sA\sD\sM-\sL}}{R - 2 \, \slLambda }}   \m . 
\label{Elim-alpha} 
\ee 
{\bf Remark 2} Substituting this into $\u{\sbcM}$ gives the {\it thin sandwich equation} \cite{BO69}
\be 
{\cal D}_{j}  \left\{ \frac{  \sqrt{  {\cal R} - 2 \, \slLambda  }  }
                           {  ||\delta_{\vec{\upbeta}} \bh ||_{\sbM}  }    
                          \{ \mh^{jk} {\delta^{l}}_{i} - {\delta^{j}}_{i} \mh^{kl} \}\{ \delta_{\vec{\upbeta}} \mh_{kl} \}  \right\}  \es  0                               \m . 
\ee
{\bf Remark 3} As an explicit PDE,  
\be
{\cal D}_{j}\left\{  \sqrt{  \frac{  {\cal R} - 2 \, \slLambda  }
                                  {  \{ \mh^{ac} \mh^{bd} - \mh^{ab} \mh^{cd} \}\{ \dot{\mh}_{ab} - 2 \, {\cal D}_{(a}\upbeta_{b)} \} 
								                                                \{ \dot{\mh}_{cd} - 2 \, {\cal D}_{(a}\upbeta_{b)} \}    }  }   
                                     \{ \mh^{jk} {\delta^{l}}_{i} - {\delta^{j}}_{i} \mh^{kl} \}\{ \dot{\mh}_{kl} - 2{\cal D}_{(k}\upbeta_{l)} \}  \right\}     \es  0    \m .  
\label{Thin-San-Eq}
\eeq
{\bf Definition 1} Solving the Lagrangian form of the GR momentum constraint $\u{\sbcM}$ -- the so-called thin sandwich equation -- 
with {\it thin sandwich data} 
\be 
( \bh, \m  \dot{\bh} ) 
\ee
for the shift $\u{\beta}$ is termed the {\it Thin Sandwich Problem} \cite{WheelerGRT, BO69, BF93} (Fig \ref{CR-Development}.b).  

\m

\n With $\u{\beta}$ found, extrinsic curvature can furthermore be computed giving a local coating of spacetime.  

\m 

\n{\bf Remark 1} `Thin Sandwich' is named with reference to the Thick Sandwich Problem \cite{WheelerGRT}, of which it is the `thin limit'. 
Here, the bounding `bread-slice' {\it thick sandwich data}
\be  
( \bh^{(1)}, \m  \bh^{(2)} )  
\ee
to solve for the GR spacetime `filling' in between. 
This was an attempted analogy with Feynman's path integral for quantum transition amplitudes between states at two different times \cite{WheelerGRT};  
this however fails to be mathematically well-posed.  

\m 

\n{\bf Remark 2} The Thin Sandwich Problem is moreover one of the Problem of Time facets, from Wheeler's account in \cite{WheelerGRT} 
                                                                                                  to Isham and \K's reviews \cite{K92, I93}.  

\m

\n{\bf Remark 3} The Thin Sandwich facet has always been presented as a manifestly classical-level problem.

\m

\n{\bf Remark 4} It is indeed a problem concerning time because, firstly, it is solving for a local slab of GR spacetime immediately adjacent to $\bupSigma$.

\m 

\n{\bf Remark 5} Secondly, the Thin Sandwich  is prerequisite for various Problem of Time strategies 
-- including our emergent time one in the presence of nontrivial Configurational Relationalism -- 
due to the GR momentum constraint $\u{\sbcM}$ interfering with resolutions of the Frozen Formalism Problem (see Part II for more).   

\m

\n{\bf Remark 6} The Thin Sandwich Problem is moreover a major mathematical PDE problem; 
see Appendix O of \cite{ABook} for an outline and \cite{BO69, BF93} for original literature.

\subsection{The Baierlein--Sharp--Wheeler (BSW) action}

{\bf Structure 1} We can also substitute (\ref{Elim-alpha}) into the ADM action.  
This produces the {\it Baierlein--Sharp--Wheeler (BSW)} \cite{BSW} action
\be 
{\cal S}_{\sG\sR}^{\sB\sS\sW} \es \int \d\lambda \int_{\sbupSigma} \sqrt{\mh} \sqrt{{\cal T}_{\sG\sR}^{\sB\sS\sW} \{{\cal R} - 2 \, \slLambda \}}  \m .   
\ee
The kinetic term here is  
\be  
{\cal T}^{\sG\sR}_{\sB\sS\sW}  \:=  {||  \{ \cdot{\mbox{ }} - \pounds_{\beta}  \} \bh||_{\sbM}}^2  
                               \es  \mM^{abcd}
\left\{ 
\dot{\mh}_{ab}  - \pounds_{\u{\upbeta}} \mh_{ab} 
\right\}
\left\{ 
\dot{\mh}_{cd}  - \pounds_{\u{\upbeta}} \mh_{cd}   
\right\}                                                \m ,  
\ee 
now and henceforth using the undensitized version of $\bM$.  

\m

\n{\bf Remark 1} This is clearly an instance of Lagrange multiplier elimination.  

\m 

\n{\bf Remark 2} This equivalence is a partial analogue of the Euler--Lagrange to Jacobi equivalence of Sec I.4.5, which is rather a passage to the Routhian.  
This analogy is moreover rendered exact in Article VI's reformulation.
 
\m 
 
\n{\bf Structure 2} The BSW expression for the gravitational momenta is 
\be 
\bp  \es  \frac{  \delta {\cal L}  }{  \pa \dot{\bh}  } 
     \es  \sqrt{    \frac{  {\cal R} - 2 \, \slLambda  }{  {\cal T}_{\sG\sR}^{\sB\sS\sW}  }  } \sqrt{h} \{ \bcalK - {\cal K} \, \bh  \}  \m .
\ee
\n{\bf Proposition 5} ${\cal S}_{\sB\sS\sW}$ encodes the standard momentum-variables form of $\u{\sbcM}$. 
in terms of velocities, however, this arises directly in the form of the thin sandwich equation. 

\m 

\n{\bf Corollary 3} An action principle directly giving the thin sandwich equation is provided by the BSW action.

\subsection{The BSW action and Temporal Relationalism}

\n{\bf Remark 1} The ADM action fails Temporal Relationalism i) since the lapse, signifying `time elapsed', is an extraneous timelike variable.

\m 

\n{\bf Remark 2} While the Misner action is Manifestly Reparametrization Invariant, 
         the BSW    action is not, since the shift $\beta$ corrections breaks this property of the metric velocity terms.
The BSW action is however close enough to implementing Manifest Reparametrization Invariance that it already encodes $\scH$ as a primary constraint,  
\be 
||\bp||_{\sbN}     =  \left|\left|  \sqrt{\frac{\overline{{\cal R} - 2\Lambda}}{{\cal T}_{\sG\sR}^{\sB\sS\sW}}} \u{\u{\u{\u{\bM}}}} \cdot 
                                                                                                        \delta_{\vec{\upbeta}} \u{\u{\bh}}  \right|\right|_{\sbN} 
                   =                \sqrt{\frac{\overline{{\cal R} - 2\Lambda}}{{\cal T}_{\sG\sR}^{\sB\sS\sW}}}        || \delta_{\vec{\upbeta}} \bh              ||_{\sbM\sbN\sbM} 
                   =                \sqrt{\frac{\overline{{\cal R} - 2\Lambda}}{{\cal T}_{\sG\sR}^{\sB\sS\sW}}}        || \delta_{\vec{\upbeta}} \bh              ||_{\sbM}   
                   =                \sqrt{\frac{\overline{{\cal R} - 2\Lambda}}{{\cal T}_{\sG\sR}^{\sB\sS\sW}}}         \sqrt{ {\cal T}_{\sG\sR}^{\sB\sS\sW}}
				   =                \sqrt{      \overline{{\cal R} - 2\Lambda}    }                                                                                    \m . \m \Box
\ee 
Overline here denotes densitization. 
In Article VI, moreover, we find actual Manifest Reparametrization Invariant and 
                                        Manifest Parametrization Invariant--dual-$\FrQ$-Geometry actions for GR.  
While this is `Machianized' by further development of actions (also in Article VI), 
the PDE nature of the elimination is moreover invariant under these developments.  

\m 

\n The Thin Sandwich is sequentially generalized by the following two sections' notions, as per Fig \ref{CR-Development}'s end summary.

\section{Best Matching (intermediate-level facet 0b)}\label{BM}

\subsection{Introducing the Relational Particle Mechanics (RPM) arena}

There was historically a lack in viable relational theories or formulations of Mechanics.  
{\it Relational Particle Mechanics (RPM)} theories are comparably recent, starting with that of Barbour and Bertotti (1982) \cite{BB82}.  
This is based on Euclidean transformations.  
RPMs are the second main model arena used in this Series of Articles; 
on occasion we make use of the similarity shapes version \cite{Kendall84, B03} of this as well.  

\m 

\n{\bf Model 1} Scaled relational particle configurations consist of just relative angles and relative separations. 
A model arena in which just these are meaningful is {\it Euclidean Relational Particle Mechanics (RPM)} \cite{BB82, FileR, QuadI}.  

\m 

\n{\bf Model 2} On the other hand, `pure-shape' relational particle configurations involve just relative angles and {\sl ratios} of relative separations.  
{\it Similarity RPM} \cite{B03, FORD, FileR} is a model arena in which just these are meaningful.

\m 

\n Sec I.2's constellationspace $\FrQ(d, N) = \mathbb{R}^{dN}$ of $N$ particles is a redundant configuration space for these theories.  
This possesses the obvious $\mathbb{R}^{dN}$ Euclidean metric. 

\m 

\n{\bf Motivation} The minisuperspace model arena subcase is of no use for Configurational Relationalism, since this facet is trivial therein.  
On the other hand, RPM is a useful model arena by, complementarily, exhibiting Configurational Relationalism, 
and notions of structure and thus of structure formation, which are also absent from Minisuperspace.
For, in RPMs, linear constraints and inhomogeneities are logically-independent features. 
In contrast, in the Minisuperspace arena, both are concurrently trivialized by 
homogeneity rendering the spatial derivative operator $\bcalD$ meaningless.
Relational Particle Mechanics furthermore points to a theory of Quantum Background Independence \cite{ABook}, 
that is complementary to `Quantum Gravity' interpreted literally.

\subsection{Euclidean RPM}

\n{\bf Structure 1} Euclidean RPM can be taken to be fundamental rather than effective Mechanics, by which it makes sense for the corresponding potentials to be of the form \cite{FileR}
\be 
V(\biq)  \es  V\left(\underline{q}^I \cdot \underline{q}^J  \mbox{ alone}\right)  \m .
\ee 
This form then guarantees that auxiliary translation and rotation corrections applied to this part of the action straightforwardly cancel each other out within.  

\m  

\n{\bf Structure 2} The kinetic term is more complicated, because $\d/\d\lambda$ is not a tensorial operation under the $\lambda${\it -dependent Euclidean group}, 
\be
T_{\sE\sR\sP\sM}                                             \es  \half ||\Circ_{\underline{A}, \underline{B}}\biq||_{\sbim}\m^2  \m , \m \mbox{ for }  \m   
\Circ_{\underline{A}, \underline{B}} \, \biq  \:=  \dot{\biq} - \underline{A} - {\underline{B}} \cr \biq                                 \m 
\label{T-ERPM}
\ee
the `$Eucl(d)$-{\it corrected derivative}'.

\m 

\n{\bf Structure 3} The Euclidean RPM action is then 
\be
{\cal S}_{\sE\sR\sP\sM}  \es   2 \int \d\lambda \sqrt{W \, T_{\sE\sR\sP\sM}}                                                                                         \m .
\label{ERPM-Ac} 
\ee
\n{\bf Structure 4} The momenta conjugate to the $\biq$ are 
\be 
p_I  \es  \sqrt{ \frac{W}{T} } \, m_I \delta_{IJ} \Circ_{\underline{A}, \underline{B}} \, q^J                              \m .
\ee
\n{\bf Structure 5} By virtue of Manifest Reparametrization Invariance and the particular square-root form of the Lagrangian, 
these momenta obey a primary constraint that is purely quadratic in the momenta ($\bn := \bm^{-1}$), 
\be
\scE  \:=  \half {||\bip||_{\sbin}}^2 + V(\biq)  
      \es  E                                                  \m . 
\label{calE}
\ee
\n{\bf Structure 6}  Variation with respect to $\underline{A}$ and $\underline{B}$ give secondary constraints, respectively,    
\be
\u{\sbcP}  \:=  \sumIN \underline{p}_{I}  
                  \es  0 \m \m \mbox{ (zero total momentum constraint) }                  \mma 
\label{ZM}       
\ee
\be
\u{\sbcL}  \:=  \sumIN \underline{q}^{I} \cr \underline{p}_{I} 
                  \es   0 \m \m \mbox{ (zero total angular momentum constraint) }         \m . 
\label{ZAM}                                                                                                           
\ee
\n{\bf Remark 1} $\u{\sbcP}$ and $\u{\sbcL}$ are homogeneous-linear in the momenta.  

\m 

\n{\bf Remark 2} $\u{\sbcP}$ can furthermore be interpreted as passing to, or factoring out, centre of mass motion. 
All the tangible physics resides in the remaining relative vectors between particles.     

\m

\n{\bf Remark 3} Returning to the Best Matching procedure, the constraints (\ref{ZM}, \ref{ZAM}), rewritten in Lagrangian configuration--velocity variables  
are to be solved for the auxiliary variables $\underline{A}$, $\underline{B}$ themselves.
This solution is then substituted back into the action, so as to produce a final $Tr$- and $Rot$-independent expression that {\sl directly} implements Configurational Relationalism.  
One has the good fortune of being able to solve Best Matching explicitly for a wide range of RPMs; see Sec \ref{RPM-BM} for a few details.

\subsection{Similarity RPM}

{\bf Structure 1} This is most naturally formulated with potentials of the form \cite{FileR}
\be 
V(\biq) = V \left( \frac{\u{q}^I \cdot \u{q}^J}{\u{q}^K \cdot \u{q}^L  } \mbox{ alone} \right)  \m .
\ee 
\n{\bf Structure 2} This form then guarantees that auxiliary translation, rotation and scale corrections applied to this part of the action 
straightforwardly cancel each other out within.  
The kinetic term is, once again, more complicated, because $\d/\d\lambda$ is not a tensorial operation under the $\lambda${\it -dependent similarity group}, giving   
\be
T_{\sS\sR\sP\sM}                                             \es  \frac{1}{2 \, I} ||\Circ_{\underline{A}, \underline{B}, C}\biq||_{\sbim}\m^2  \m , \m \mbox{ for }  \m   
\Circ_{\underline{A}, \underline{B}, C} \, \biq  \:=  \dot{\biq} - \underline{A} - {\underline{B}} \cr \biq + C \biq                                 \m 
\label{T}
\ee
the `$Sim(d)$-{\it corrected derivative}'.

\m 

\n{\bf Structure 3} The action is then 
\be
S_{\sS\sR\sP\sM}  \es   2 \int \d\lambda \sqrt{W \, T_{\sS\sR\sP\sM} }                                                                                        \m .
\label{BB-Ac} 
\ee
\n{\bf Structure 4} The conjugate momenta are 
\be 
p_I  \es  \sqrt{ \frac{W}{T_{\sS\sR\sP\sM}} } \, \frac{m_I}{I} \delta_{IJ} \Circ_{\underline{A}, \underline{B}, C} \, q^J   \m 
                                  \m .
\ee
\n{\bf Structure 5} By virtue of Manifest Reparametrization Invariance and the particular square-root form of the Lagrangian, 
these momenta obey a primary constraint that is purely quadratic in the momenta, 
\be
\scE  \:=  \frac{1}{2 \, I} {||\bip||_{\sbin}}^2 + V(\biq)  
      \es  E                                                  \m . 
\label{calE2}
\ee
{\bf Structure 6} Variation with respect to $\underline{A}$, $\underline{B}$ and give secondary constraints (\ref{ZM}, \ref{ZAM}) again and 
\be
\scD  \:=  \sumIN \u{q}^I \cdot \u{p}_I  
                  \es  0 \m \m \mbox{ (zero total dilational momentum constraint) }  \m . 
\label{ZDM}     
\ee
\n{\bf Remark 1} This is also homogeneous-linear in the momenta.

\subsection{The general Manifestly Reparametrization Invariant case of Best Matching}

Let us now formulate it for a {\it general} theory at the differential-geometric level of structure. 

\m 

\n{\bf Best Matching 0} $\lFrg$-correct the action's velocities according to the infinitesimal Lie-derivative (Article XIV) action 
\be 
\dot{\biQ} \m \longrightarrow \m \Circ_{\sbig} \biQ   \es  \dot{\biQ} - \pounds_{\sbig}\dot{\biQ}   \m .  
\label{Lie-Drag-1}
\ee
$\lFrg$ is to act on $\FrQ$ as a {\sl shuffling group}, in the sense that infinitesimally separated pairs of configurations are considered, 
with one kept fixed and the other shuffled around.  
$\bigg$ are $\lFrg$-auxiliary variables, indexed by $\fG$.  

\m 
 
\n{\bf Best Matching 1} We vary our action with respect to $\bigg$ to obtain secondary constraints 
\be 
\bShuffle = 0 \m ,  
\ee
indexed likewise. 

\m 

\n {\bf Remark 1} Configurational Relationalism is thus a Constraint Provider. 

\m 

\n{\bf Remark 2} Since the $\bShuffle$ arise by variation, they are by definition secondary.  

\m 

\n Being linear in the momenta, they could also be denoted by $\bLin$.
These are however but a subcase of the most general linear constraints $\bLin$ (indexed by $\fL$ whose range is in general distinct from $\fG$), 
since not all possible such may arise from shuffling.

\m 

\n For now, we additionally assume that the $\bShuffle$ are first-class, and are a fortiori gauge constraints. 
It is important to note that, while these may be expectation to have, or are all that a reader is so far familiar with, 
these are in fact matters to be checked.  
This is a matter of Constraint Closure as covered in Articles III and VII, with \cite{HTBook} as a notable preliminary review.  
We only change notation to $\bFlin$ and to $\bGauge$ when the corresponding checks have been confirmed.  

\m 

\n{\bf Remark 3} The initial introduction of $\lFrg$ corrections appears at first sight to be a step in the wrong direction 
as regards freeing Physics on the configuration space $\FrQ$ from the meaningless transformations $\lFrg$.  
For it extends the already redundant space $\FrQ$ of the $\biQ$ to some joint space of $\biQ$ and the $\lFrg$-auxiliary variables $\bigg$.  
If $\bShuffle$ does turn out to be of the form $\bFlin$, however,   
Sec III.2 explains these to be a type of constraint which uses up {\sl two} degrees of freedom per $\lFrg$ degree of freedom.  
Each degree of freedom appended then wipes out not only itself but also one of $\FrQ$'s redundancies. 
Thus indeed one ends up on a $\FrQ$ that is free of these redundancies -- the quotient space $\w{\FrQ} = \FrQ/\lFrg$ -- 
as is required to successfully implement Configurational Relationalism.  

\m 
												  
\n{\bf Best Matching 2)} Next solve our general classical theory's $\bShuffle$ 
in Lagrangian formulation -- the so-called {\it best-matching equations} --
\be 
\bFlin(\biQ, \dot{\biQ}, \bigg) = 0 \m , 
\ee
with {\it best matching data} 
\be 
( \biQ , \m  \dot{\biQ} ) 
\ee
for the $\lFrg$-auxiliaries $\bigg$.  
I.e.\ solve the Lagrangian form of the constraints arising from variation with respect to the $\lFrg$-auxiliaries $\bigg$ 
                                                                                      for the $\lFrg$-auxiliaries $\bigg$ themselves.
This is termed the {\it Best Matching Problem} \cite{FileR, APoT2} (Fig \ref{CR-Development}.c).  

\m 

\n{\bf Remark 4} Solving this amounts to shuffling into extremal incongruence 
(it is only minimal incongruence if established by further standard Calculus workings).

\m 

\n{\bf Best Matching 3)} Next substitute these `best-matched' extremizing solutions back in the original Principles of Dynamics action 
to obtain a reduced action, i.e.\ on the reduced configuration space $\w{\FrQ}$.  
This amounts to completing the Lagrange multiplier elimination of the $\bigg$.

\m 

\n{\bf Best Matching 4)} This resultant Principles of Dynamics action is finally elevated to constitute a new starting point.

\subsection{Best Matching in RPM}\label{RPM-BM}

{\bf Structure 1} Here $\lFrg = Eucl(d)$ or $Sim(d)$ is the physically-irrelevant group of automorphisms acting on the ab initio configuration space 
$\FrQ(d, N) = \mathbb{R}^{N \, d}$: constellationspace. 

\m 

\n{\bf Remark 1} Eliminating translations $Tr(d)$ is very straightforward. 
It can be envisaged as passing to the centre of mass, or as declaring the position of the centre of mass of the universe as a whole to be meaningless. 
One can moreover rewrite this action to look just like the absolute action except with one particle label less. 
This is attained by diagonalization to Jacobi coordinates; it helps to continue working with these as per below.  
The corresponding action is now on relative space, i.e.\ constellationspace quotiented by translations, which is just $\mathbb{R}^{n \, d}$ for $n := N - 1$. 

\m 

\n{\bf Remark 2} Treating similarity RPM is in some ways simpler \cite{FORD}; in particular, eliminating dilations $Dil$ is straightforward as well.  
This amounts to passing to the obvious unit sphere $\mathbb{S}^{n \, d - 1}$ within relative space.  

\m 

\n{\bf Proposition 6} In 1-$d$, there are no rotations, so the above workings already suffice for Euclidean and similarity RPMs in this case.  

\m 

\n In 2-$d$, further eliminating the rotations $Rot(d)$ gives \cite{FORD} the Fubini--Study metric on complex-projective space $\mathbb{CP}^{N - 2}$ 
as similarity RPM's final reduced configuration space: shape space.  

\m

\n The $N = 3$ particles case is furthermore special \cite{FORD} 
since $\mathbb{CP}^1 = \mathbb{S}^2$ and the Fubini--Study metric reduces to the standard spherical metric as well.  

\m 

\n{\bf Proposition 7} For Euclidean RPM, dilations are either never removed or are `coned back in', with the corresponding reduced configuration space 
-- relational space -- being \cite{FileR} the topological and metric cone over the corresponding shape space. 
For $N = 3$ in 2-$d$, this returns $\mathbb{R}^3$ albeit with a curved rather than flat metric upon it.  

\m 

\n{\bf Remark 3} The above structures and examples are further defined, detailed and discussed in Sec \ref{Direct}, 
given that they coincide with what a direct study of spaces of (scaled) shapes find. 

\m 

\n{\bf Remark 4} See Article V for how this working can be extended to obtain RPM's version of emergent Machian time.

\subsection{GR as an already-best-matched theory}

{\bf Structure 1} In GR-as-Geometrodynamics, the spatial diffeomorphisms 
\be
\lFrg = Diff(\bupSigma)
\ee 
play the role of automorphism group acting on the incipient configurations regarded to be physically meaningless.
These ara analogous to considering $Rot(d)$ acting on relative space, including in having scale, though it is now a {\sl local} notion of scale.
These are moreover now obligatory rather than optional, for the reason given in Sec III.2. 

\m 

\n Modelling this is a Differential Geometry level endeavour, implemented infinitesimally by the Lie derivative: 
on this occasion a {\sl well-known} application thereof.    
In the usual case of 3 spatial dimensions, the spatial diffeomorphisms use up 3 degrees of freedom per space point.  

\m 

\n{\bf Remark 1} Inspecting the BSW action, this arises in already -$Diff(\bupSigma)$-corrected, 
with the corresponding $\u{\beta}$ variation and the substitution of resultant extremal value back into the action constituting Best Matching.  

\m 

\n{\bf Best Matching 1} Introduce GR'as-Geometrodynamics' $Diff(\bupSigma)$-corrected derivative
\be 
\Circ_{\u{\upbeta}} \bh   \m .  
\ee 
This is {\sl computationally} the same as the hypersurface derivative 
\be 
\delta_{\vec{\upbeta}} \bh                                           \m , 
\ee 
but is {\sl conceptually} and even {\sl structurally} different. 
For the former is a spatial-primality concept while the latter is a spacetime-primality structure, thus presupposing further structure.  

\m 

\n{\bf Aside 1} We thus now view the GR kinetic term as
\be 
{\cal T}_{\sG\sR}^{\sB\sM}  \es  {||\Circ_{\u{\upbeta}} \bh ||_{\sbM}}^2      \m . 
\ee 
$\scH$ now arises as a primary constraint according to 
\be 
||\bp||_{\sbN}    =  \left|\left|  \sqrt{\frac{\overline{{\cal R} - 2\Lambda}}{{\cal T}_{\sG\sR}^{\sB\sM}}} \u{\u{\u{\u{\bM}}}} \cdot 
                                                                                           \Circ_{\u{\upbeta}} \u{\u{\bh}}  \right|\right|_{\sbN} 
                  =                \sqrt{\frac{\overline{{\cal R} - 2\Lambda}}{{\cal T}_{\sG\sR}^{\sB\sM}}}        || \Circ_{\u{\upbeta}} \bh              ||_{\sbM\sbN\sbM} 
                  =                \sqrt{\frac{\overline{{\cal R} - 2\Lambda}}{{\cal T}_{\sG\sR}^{\sB\sM}}}        || \Circ_{\u{\upbeta}} \bh              ||_{\sbM}   
                  =                \sqrt{\frac{\overline{{\cal R} - 2\Lambda}}{{\cal T}_{\sG\sR}^{\sB\sM}}}         \sqrt{ {\cal T}_{\sG\sR}^{\sB\sM}}
				  =                \sqrt{      \overline{{\cal R} - 2\Lambda}      }                                                                                 \m . \m \Box
\ee
\n{\bf Best Matching 2)} gives $\bShuffle = \u{\sbcM}$ -- the GR momentum constraint -- in the usual way, 
whereas {\bf Best Matching 3)} is just the Thin Sandwich treatment of $\u{\sbcM}$.   

\m 

\n{\bf Remark 2} See Article VI for how this working can be extended to obtain GR's version of emergent Machian time.
%
%
In this way, GR does happen to implement \cite{B94I, RWR, AM13} 
the philosophically desirable nugget that is Mach's Time Principle can be implemented for Einstein's GR.  
%
%
See there also for Best Matching in Electromagnetism, Yang--Mills Theory and for each of these coupled to GR.

\section{Configurational Relationalism}\label{Intro-CR}

\subsection{Underlying Principles of Leibniz and Mach}\label{LMP}

The current Article's aspect of Background Independence is grounded on the following. 

\m 

\n{\bf Leibniz's Space Principle} is that {\sl space is the order of coexisting things} \cite{L}.

\m 

\n{\bf Mach's Space Principle} is that \cite{M} {\it ``No one is competent to predicate things about absolute space and absolute motion. 
These are pure things of thought, pure mental constructs that cannot be produced in experience. 
All our principles of mechanics are, as we have shown in detail, experimental knowledge concerning the relative positions of motions and bodies."} 

\m  

\n{\bf Remark 1}  This is not to be confused with Mach's Principle for the Origin of Inertia (\cite{Buckets}, Section 3 of \cite{ABook}).
Or with anything else called Mach's Principle by some author or other (e.g.\ \cite{Buckets, RovelliBook} having an extensive selection of other such uses).

\subsection{Configurational Relationalism's postulates}

This Background Independence aspect's true-name is {\bf Configurational Relationalism}. 
It is moreover the {\sl first} aspect to consider in building our local Problem of Time resolution, by which this resolution can be said to have grown from Shape Theory...

\m

\n a) {\bf Spatial Relationalism} \cite{BB82} is to not ascribe any absolute properties to space.

\m 

\n b) {\bf Internal Relationalism} is the post-Machian addition of also not ascribing any absolute properties to any additional internal space associated with the matter fields. 
This is both a useful addition and straightforward.

\m

\n Internal Relationalism is substantially distinct though holding at a fixed spatial point whereas a) moves spatial points around.  
Configurational Relationalism is then approached as follows.

\m 

\n{\bf Configurational Relationalism i)} One is to include no extraneous configurational structures 
(spatial or internal-spatial metric geometry variables of a fixed-background rather than dynamical nature) \cite{FileR}.

\m 

\n{\bf Configurational Relationalism ii)} Physics in general involves not only a $\FrQ$ 
but also a group $\lFrg$ of transformations acting upon $\FrQ$ that are taken to be physically redundant \cite{FileR}.

\m 

\n{\bf Remark 1} Since time-parametrization really involves a 1-$d$ metric of time, Temporal Relationalism i) and 
                                                                Configurational Relationalism i) 
reflect a single underlying relational conception of Physics: that there is to be no fixed-background Metric Geometry.  

\m

\n{\bf Remark 2} ii) is a matter of practical convenience: often $\FrQ$ with redundancies is simpler to envisage and calculate with.
The Internal Relationalism case of Configurational Relationalism ii) is a distinct formulation of Gauge Theory (as per Article VII) 
from the conventional one presented in Article X.   
The spatial case is similar: it can also be thought of as a type of Gauge Theory for space itself.\footnote{The name and concept of Gauge Theory 
is used here in a somewhat broader manner than that of Particle Physics, covering also e.g.\ Molecular Physics \cite{LR97} and cosmological perturbations \cite{CP}.}    
%
This includes modelling translations and rotations relative to absolute space as redundant in Mechanics, for which $\FrQ = \mathbb{R}^{dN}$, 
                                                         or $Diff(\bupSigma)$ as redundant in GR,        for which 

\n	                                                           													   $\FrQ = \Riem(\bupSigma)$.
%
%
Section \ref{Gact-Gall} subsequently discuss restrictions on $\FrQ$, $\lFrg$ pairings.  

\m 

\n Two a priori distinct conceptualizations of Configurational Relationalism in the point particle setting are as follows.  

\m 

\n{\bf Structure Acted Upon a)} $\lFrg$ acts on absolute space $\Abs(d)$ (usually taken to be $\mathbb{R}^d$).  

\m 

\n{\bf Structure Acted Upon b)} $\lFrg$ acts on configuration space $\FrQ(d, N)$, i.e.\ it acts, rather, on material entities of at least some physical content.

\m 

\n{\bf Remark 1} b) can be approached via a)'s groups acting on $\mathbb{R}^d$.  

\m 

\n{\bf Remark 2} a) is moreover well-known: Klein's Erlangen Program for Geometry. 
We have thus converted an `a priori new' problem in Relational Mechanics into an issue already long considered in Geometry.  

\m 

\n Some useful limitations on the choice of $(\lFrg, \FrQ)$ pairs are as follows.

\m 

\n{\bf Criterion C)} {\it Nontriviality}. $\lFrg$ cannot be too large, i.e. a bounding degrees of freedom count criterion. 
Using $k := \mbox{dim}(\FrQ)$     and $l :=  \mbox{dim($\lFrg$)}$, a theory on $\FrQ/\lFrg$ is 
\be 
\mbox{{\it inconsistent} if } \m           l > k       \m ,
\ee 
\be 
\mbox{{\it utterly trivial} if } \m        l = k \mma \mbox{ or } 
\ee 
\be 
\mbox{{\it relationally trivial} if } \m  l = k - 1   \m . 
\ee 
\n{\bf Remark 3} Relational nontriviality is meant here in the sense of requiring at least two degrees of freedom, 
so that at least one of these can be expressed in terms of at least one other such.
This is to be contrasted with the idea of degrees of freedom being `meaningfully expressed' in terms of some external or otherwise unphysical `time parameter'.   

\m 
 
\n{\bf Criterion B)} Further {\it structural compatibility} is required.  
A simple example of this is that if one is considering $d$-dimensional particle configurations, then $\lFrg$ is to involve the same $d$ (or smaller, but certainly not larger).  

\m 

\n{\bf Criterion A)} A more general structural compatibility criterion is for $\lFrg$ is to admit a group action on $\FrQ$. 
A group action's credibility may further be enhanced though its being `natural'.
Some further mathematical advantages are conferred from group actions being one or both of faithful or free, 
with the combination of free and proper conferring yet further advantages. 

\m 

\n One might additionally wish to choose $\lFrg$ for a given $\FrQ$ so as to eliminate {\sl all} trace of extraneous background entities.

\m 

\n{\bf Choice 1} The automorphism group $Aut(\Abs)$ of absolute space $\Abs$ is an obvious possibility for $\lFrg$. 

\m 

\n{\bf Choice 2} Some subgroup of $Aut(\Abs)$  \cite{Kobayashi} might however also be desirable, 
for instance because the inclusion of some automorphisms depends on which level of mathematical structure $\sigma$ is to be taken to be physically realized.
In this way, 
\be 
\lFrg \m \leq \m  Aut(\langle \Abs, \sigma \rangle)
\ee 
for some $\sigma$ is a more general possibility.
%

\m 

\n{\bf Remark 4} The next two sections cover one distinct implementation of Configurational Relationalism each.  
Such seeking can be either  indirect by applying a correctory                  \cite{BB82, B03}     group action on unreduced configuration spaces 
                  or     by direct formulation on reduced configuration spaces \cite{FORD, FileR}: `relational spaces'.

\section{Direct implementation}\label{Direct} 

\n A fortunate few cases admit a direct formulation: we can write down a Mechanics action directly on $\w{\FrQ}$ \cite{FORD, FileR}.  
We call this a {\it relational action}.
\cite{FileR} moreover demonstrated that reduced and relational coincide for similarity and Euclidean RPMs.
By this, the joint-summary name {\sl r-formulation} is appropriate in cases of confluence.  
In this case, one can work solely with $\lFrg$-invariant objects $O(\w{\biQ})$.  

\m 

\n{\bf Structure 1} Given a candidate pair $(\FrQ$, $\lFrg)$, one seeks to represent the generators of $\lFrg$ (indexed by $\fG$) as 
\be
\left( \biQ, \, \frac{\pa}{\pa \sbiQ} \right)
\ee 
which manifestly act on $\FrQ$,   

\m 

\n{\bf Remark 1} Knowing what a particular $\w{\FrQ}$'s geometry is, 
and this so happening to be whichever of simple or already well studied, may be permissive of a direct approach.
Much of Kendall's work \cite{Kendall} on Similarity Shape Theory benefits from this being the case in 2-$d$ 
(and in 1-$d$, though this case was already known prior to Kendall).

\subsection{Totally trivial examples: $\lFrg = id$}

\n 1) Article I's spatially-absolute Mechanics. 

\m 

\n 2) Minisuperspace, including e.g.\ Article I's specific minisuperspace models.

\subsection{Geometrically trivial examples}

\n{\bf Example 1)} Translation-invariant Mechanics, for which the reduced configuration space is {\it relative space}: 
\be 
\lFrr(d, N)  \es  \frac{\FrQ(d, N)}{\mbox{Tr}(d)}  
             \es  \mathbb{R}^{nd}
\ee			 
for $n := N - 1$.

\m 

\n{\bf Remark 1} In the Newtonian Paradigm, this amounts to passing to the centre of mass frame as a matter of convenience.
In the Relational Approach, on the other hand, the centre of mass position for the whole Universe is a fortiori meaningless.
In either case, setting the more usual theories of Mechanics free from overall translations is trivial.  

\m 

\n{\bf Remark 2} See Fig \ref{Relative-Coordinates} for various useful coordinate systems for $\lFrr(d, N)$.  
%
{            \begin{figure}[!ht]
\centering
\includegraphics[width=0.72\textwidth]{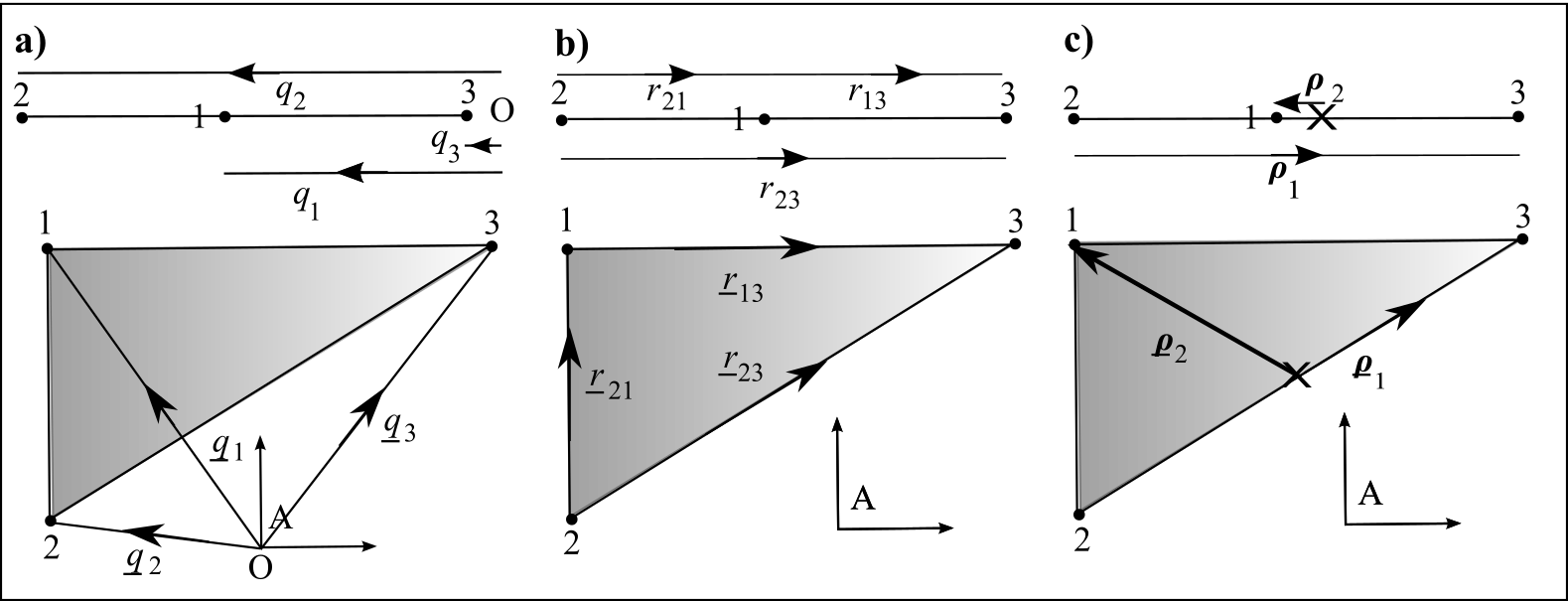}
\caption[Text der im Bilderverzeichnis auftaucht]{        \footnotesize{Coordinate systems for 3 particles in each of 1- and 2-$d$. 

\m 

\n a)--b) Absolute particle position coordinates ($\uiq_1$, $\uiq_2$, $\uiq_3$).  
These are defined with respect to fixed absolute origin O and axes A.  

\m 

\n c)--d) Relative inter-particle (Lagrange) coordinates $\bir := \{\uir^{IJ}, I > J\}$.
Their relation to the $\uiq^I$ is obvious: eq (\ref{Lag}). 
In the case of 3 particles, any 2 of these form a basis; we use upper-case Latin indices $A, B, C$ for a basis of relative separation labels 1 to $n$.  
No absolute origin enters their definition, but reference is still made to fixed coordinate axes A.

\m 

\n e--f) Relative particle inter-cluster mass-weighted Jacobi coordinates $\birho$, which are more convenient but still involve A. 
$\mbox{\large $\times$}$ denotes the centre of mass of particles 2 and 3.} }
\label{Relative-Coordinates}\end{figure}            }

\m 

\n{\bf Structure 1} {\it Relative Lagrange coordinates}: some basis of relative inter-particle separation vectors 
\be 
\u{r}^{IJ} := \u{q}^J - \u{q}^I   \m ,
\label{Lag}
\ee
are conceptually simple natural coordinates for this.   
Fig \ref{Relative-Coordinates}.c)--d) illustrate these for 3 particles in 1- and 2-$d$, though this notion indeed trivially extends to arbitrary $N$ and $d$.  

\m 

\n{\bf Structure 2} Fig \ref{Relative-Coordinates}.e)--f)'s {\it relative Jacobi coordinates} are more mathematically convenient to work with.  
These are sets of $n$ inter-particle (cluster) separations chosen such that the kinetic term (or the corresponding arc element) is diagonal.

\m 
 
\n{\bf Remark 3} They are widely used in Celestial Mechanics \cite{Marchal} and Molecular Physics \cite{LR97}.

\m 

\n{\bf Structure 3} The diagonal form for the kinetic matrix in relative Jacobi coordinates is 
\be 
\mu_{ij AB}  \:=  \mu_A\delta_{ij}\delta_{AB}             \m ,
\ee  
for $\mu_A$ the corresponding Jacobi inter-particle cluster reduced masses $\mu_A$.  

\m 

\n{\bf Example 1} For the $N = 3$ body problem, the relative Jacobi coordinates are 
\beq
\uiR_1 = \uiq_3 - \uiq_2                                   \m  \mbox{ and } \m  
\uiR_2 = \uiq_1 - \frac{m_2\uiq_2 + m_3\uiq_3}{m_2 + m_3}                        \m  
\ee 
with corresponding Jacobi masses  
\beq
\mu_1  \es  \frac{m_2m_3}{m_2 + m_3}                      \m \mbox{ and } \m 
\mu_2  \es  \frac{m_1\{m_2 + m_3\}}{m_1 + m_2 + m_3}      \m . 
\eeq
\n{\bf Structure 4} We furthermore pass to {\it mass-weighted relative Jacobi coordinates} 
\be 
\rho^{iA} := \sqrt{\mu_A}R^{i A}                          \m .
\ee   
The kinetic metric is now just an identity array with components 
\be 
\delta_{ij}\delta_{AB}                                    \m .
\ee
The kinetic arc element is here \cite{FileR} 
\be 
\d s  =  \sqrt{ \d \rho^A \d \rho^A }                     \m .  
\ee

\subsection{Preshape space}

Kendall's Shape Theory \cite{Kendall84, Kendall89, Kendall} (see also \cite{Small, JM00, Bhatta, DM16, PE16} for reviews)
is a trove of reduced configuration space geometry work greatly advancing Relational Mechanics (see \cite{FileR, ABook, S-I, S-II, S-III, Minimal-N}). 

\m 

\n{\bf Structure 1} If absolute scale is also to have no meaning, the configuration space is Kendall's {\it preshape space} \cite{Kendall} 
\be 
\FrP(d, N)  \:=  \frac{\lFrr(d, N)}{\mbox{$Dil$}}  
            \es  \mathbb{S}^{n \, d - 1}                                                             \m ,
\ee 
of dimension $n \, d - 1$.                                                      

\m 

\n{\bf Structure 2} The {\it normalized mass-weighted relative Jacobi coordinates} are 
\beq
n^{iA}  \:=  \frac{\rho^{iA}}{\rho}  
        \es  \frac{\rho^{iA}}{\sqrt{I}} \m , 
\label{n-def}
\eeq 
where $I$    is the moment of inertia 
and   $\rho$ is the {\it configuration space radius} (alias {\it hyperradius} \cite{MFII} in the Molecular Physics literature).  

\m 

\n{\bf Structure 3} Preshape space naturally carries \cite{Kendall84} the {\it hyperspherical metric} 
\beq
\d s^2 = \sumdn \prod\mbox{}_{\mbox{}_{\mbox{\scriptsize $m$ = 1}}}^{p - 1} \mbox{sin}^2\theta_m \d\theta_p^2  \m .
\label{HS}
\eeq
The $\theta_{\barp}$ coordinates here are related to ratios of the $\rho_A$ in the usual manner 
in which hyperspherical coordinates are related to Cartesian ones \cite{FileR}.

\subsection{Relational space and shape space}

\n{\bf Structure 1} If instead absolute axes are to have no meaning, then the configuration space is  
\beq
\mbox{\it relational space }  \m \FrR(d, N)  \:=  \frac{\FrQ(d, N)}{Eucl(d)}  
                                             \es  \frac{\Frr(d, N)}{\mbox{$Rot$($d$)} } \m .  
\eeq
\n{\bf Structure 2} If neither absolute axes and absolute scale are to have no meaning, then the configuration space is Kendall's \cite{Kendall} 
\beq
\mbox{\it shape space } \m \FrS(d, N)  \:=  \frac{\FrQ(d, N)}{\mbox{Sim}(d)}  
                                       \es  \frac{\FrP(d, N)}{\mbox{Rot}(d)}                          \m . 
\eeq
Since the dimension of this plays a recurring role in this Series of Articles, we give a notation for it,  
\beq
k(d, N)   \:=  \mbox{dim}(\FrS(d, N))  
          \es  d \, N - \frac{d\{d + 1\}}{2} + 1  
	   	  \es  \frac{d\{2 \, n + 1 - d\}}{2}   - 1                                                   \m . 
\label{qNd}
\eeq
%
%
\n{\bf Remark 1} Similarity RPM is a theory of shape, while Euclidean RPM is a theory of shape-and-scale.

\m 

\n{\bf Remark 2} The above quotient spaces are moreover taken to carry topological, differential-geometric and metric structure. 
Their analogy with GR's configuration spaces is explained in Fig \ref{Q-RPM-GR}. 

\m 

\n{\bf Remark 3} $\FrP(N, 1) = \FrS(N, 1)$, since there are no rotations in 1-$d$.  

\m 

\n{\bf Proposition 8} \cite{Kendall84, Kendall}
\be 
\FrS(N, 2) = \mathbb{CP}^{N - 2}
\ee 
with canonical Fubini--Study metric
\beq
\d s^2 = 
\frac{\{\{1 + ||\mbox{\boldmath$Z$}||_{\sC}^2\} ||\d\mbox{\boldmath$Z$}||_{\sC}^2 -  |(\mbox{\boldmath$Z$} \cdot \d \mbox{\boldmath$Z$})_{\sC}|^2\}} 
     {  \{1 + ||\mbox{\boldmath$Z$}||_{\sC}^2\}^2} 
\label{FS}
\eeq
where the $\sC$ suffix denotes the $\mathbb{C}^{n - 1}$ version of inner product and norm, 
with the {\it inhomogeneous coordinates} $Z_{\bar{p}}$'s indices running over $n - 1$ copies of $\mathbb{C}$.

\m 

\n{\bf Structure 3} Let us next introduce a generalized notion of {\it cone} over some topological manifold $\FrM$. 
This is denoted by $\mC(\FrM)$ and takes the form 
\beq
\mC(\FrM) = \FrM \times [0, \infty)/\m \widetilde{\m}  \m . 
\eeq
$\widetilde{\m}$ \m here signifies that all points of the form \{p $\in$ \FrM, 0 $\in [0, \infty)$\} 
are `squashed' or identified to a single point termed the {\it cone point}, 0. 

\m 

\n{\bf Proposition 9}  At the metric level, given a manifold $\FrM$ with a metric with line element $\d s$, 
the corresponding cone has a natural metric of the form 
\beq
\d s^2_{\scc\so\sn\se} := \d \rho^2 + \rho^2 \d s^2 \m , \m \mbox{ (for } \m \rho \in [0, \infty) \m \mbox{ a `radial' coordinate) }  \m .
\eeq
Relational space is just the cone over shape space \cite{FileR}.
This cone structure renders clear the geometrical meaning of the {\it scale--shape split} for Metric Shape and Scale RPM. 

\m 

\n{\bf Corollary 4} $\mC(\FrS(N, 1))$ is just $\mathbb{R}^{n}$ with standard flat metric.  

\m 

\n{\bf Remark 4} The coning construct is moreover independent of which shapes it is being adjoined to, 
thus constituting a `sore thumb' or heterogeneous addendum.

\subsection{Picking out the 3-stop metroland and triangleland examples}\label{Tri-Sel}

We restrict attention to Euclidean and similarity configurations and configuration spaces (see \cite{AMech, PE16, ABook, A-Killing} for more).
Let us start with the latter, due to these being geometrically simpler; furthermore they recur as subproblems within the former. 
Fig \ref{RPM-Count-Top}.a)-b) tabulates configuration space dimension $k$, so as to display inconsistency, triviality, and relational triviality by shading.
We follow this up identifying tractable topological manifolds and metric geometries in Fig \ref{RPM-Count-Top}.c)-d) \cite{FORD, ABook}.  
%
{            \begin{figure}[!ht]
\centering
\includegraphics[width=0.83\textwidth]{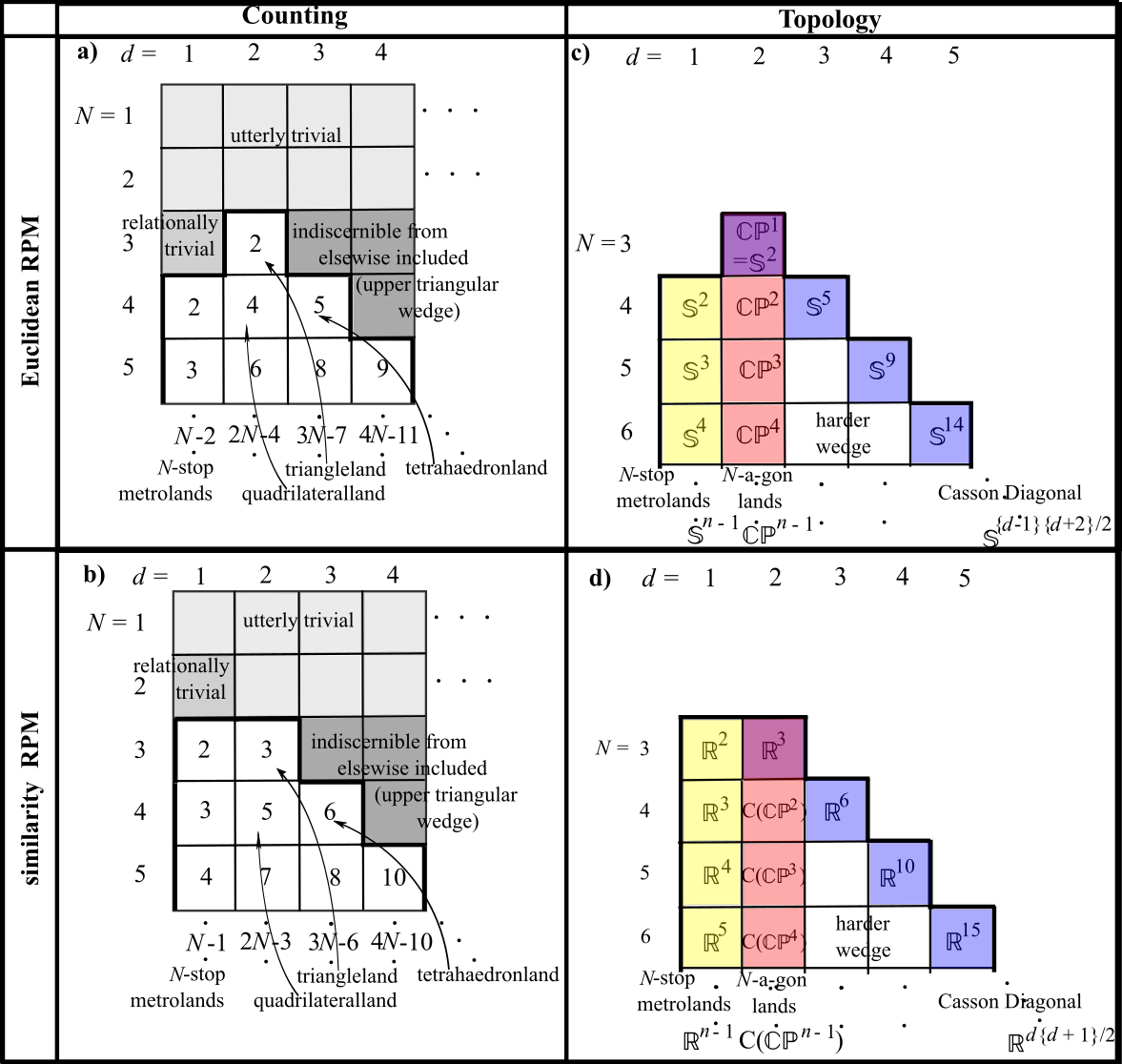}
\caption[Text der im Bilderverzeichnis auftaucht]{  \footnotesize{a) and b) are similarity and Euclidean RPM configuration space dimensions $k$ respectively.

\m 

\n c) and d) are the corresponding topological manifolds (\cite{FileR} summarizes further topological results about RPM configuration spaces).  
While this gives 3 tractable series \cite{Kendall}, only the two shaded columns are known to admit tractable metrics as well.  
Let us term 1-$d$ RPM universe models {\it N-stop metrolands} since their configurations look like underground train lines. 
We term 2-$d$ RPM universe models {\it N-a-gonlands} since their configurations are planar $N$-sided polygons. 
The mathematically highly special $N$ = 3 case of this is {\it triangleland}, 
and the first mathematically-generic $N = 4$ case is {\it quadrilateralland} \cite{QuadI}.
See \cite{Minimal-N} for more about the Casson diagonal series and tetrahaedronland.} }
\label{RPM-Count-Top}\end{figure}            }

\m 

\n{\bf Definition 1} The {\it minimal relationally nontrivial unit} \cite{AMech, ABook, Minimal-N} is concurrently the smallest relationally nontrivial 
whole-universe model, 
dynamical subsystem, and 
Shape Statistics \cite{Kendall, PE16} sampling unit.  

\m 

\n{\bf Example 1} The relational triangle (Fig \ref{CR-Development}.d) is an archetype of minimal relationally-nontrivial unit \cite{Kendall}. 
This case additionally admits some spectacular simplifications, as follows. 

\m 

\n{\bf Corollary 5} (of Proposition 8) i) In the similarity shapes case, if additionally $N = 3$, topologically 
\be 
\mathbb{CP}^1 = \mathbb{S}^2   \m .
\ee
ii) metrically, (\ref{FS}) collapses to  
\beq
\d s^2  \es  \frac{\d Z^2}{\{1 + |Z|^2\}^2}  
        \es  \d \slTheta^2 + \mbox{sin}^2\slTheta \, \d\slPhi^2  \m ,
\label{Tri-Sphericals}
\eeq
where we have used the polar form 
\be
Z  := R \, \mbox{exp}(i \, \slPhi)
\ee
where the ratio 
\be 
R \:= \frac{\rho_1}{\rho_2}
\label{Rat}
\ee 
plays the role of {\it stereographic radius} on the shape sphere, and the venerable substitution 
\be 
R = \mbox{tan}\frac{\Theta}{2}
\ee 
then takes one to the standard spherical metric.  

\m  

\n{\bf Remark 1} See Fig \ref{Relational-Coordinates} and the below two equations for the meanings of these coordinates.
%
{           \begin{figure}[!ht]
\centering
\includegraphics[width=0.55\textwidth]{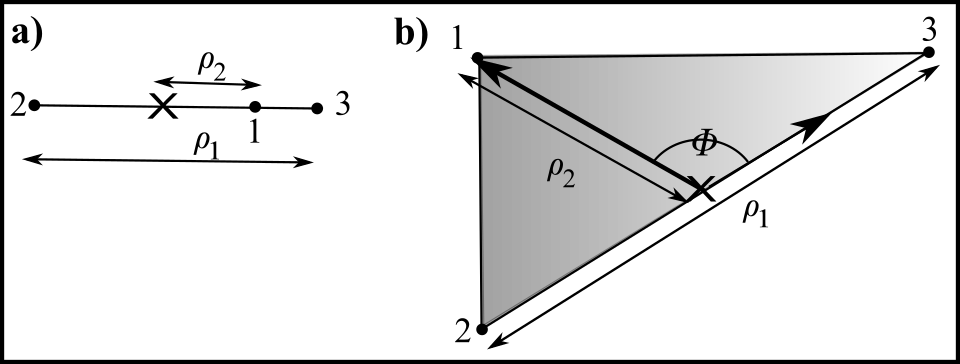}
\caption[Text der im Bilderverzeichnis auftaucht]{  \footnotesize{a) For 3 particles in 1-$d$, just use the magnitudes of the two `base' and `median' Jacobi coordinates.

\m 

\n 
b) For 3 particles in 2-$d$, use the magnitudes of the two Jacobi coordinates and define $\slPhi$ as the `Swiss army knife' angle of (\ref{Swiss}).  
This is a relative angle, so, unlike the $\brho$, these three coordinates {\sl do not} make reference to absolute axes A.  
Finally, the relative angle $\slPhi$ and some function of the ratio (\ref{Rat}) constitute pure-shape coordinates.  
In particular, (\ref{Azi}) gives the azimuth to $\slPhi$'s polar angle.} }
\label{Relational-Coordinates} \end{figure}         } 

\n Triangleland's azimuthal coordinate is the following median-to-base ratio function 
\be 
\slTheta  \:=  2\,\mbox{arctan}\left(\frac{\rho_2}{\rho_1}\right)
\label{Azi}
\ee 

\n Triangleland's polar coordinate is the following `Swiss Army knife' relative angle 
\be 
\slPhi    \:=  \mbox{arccos}\left( \frac{\brho_1 \cdot \brho_3}{\rho_1 \rho_3} \right)  \m .
\label{Swiss}
\ee 
{\bf Remark 2} Choosing triangleland as one's example `doubles' the amount of geometric and linear methods available. 
The spherical such are moreover both simpler and better-known than their complex-projective counterparts.  

\m 

\n{\bf Corollary 6} (of Proposition 9)  In the similarity shapes case, if also $N = 3$, 
\beq
\d s^2  \es  \d \rho ^2    +  \frac{\rho^2}{4}\{\d \slTheta^2 + \mbox{sin}^2\slTheta_2 \d \slPhi^2\}  
        \es  \frac{\d I^2  +  I^2\{\d \slTheta^2 + \mbox{sin}^2\slTheta \d \slPhi_2^2\}}{4I}           \m ;
\label{Tri-Scale}
\eeq
$\mC(\FrS(3, 2))$ is also $\mathbb{R}^3$, albeit not with the flat metric.  
It is, however, conformally flat \cite{Iwai87, FileR}: just apply the conformal factor $4 \, I$ to the second form of (\ref{Tri-Scale}).

\subsection{Further characterization of these examples}\label{Fur-Char}

{            \begin{figure}[!ht]
\centering
\includegraphics[width=1\textwidth]{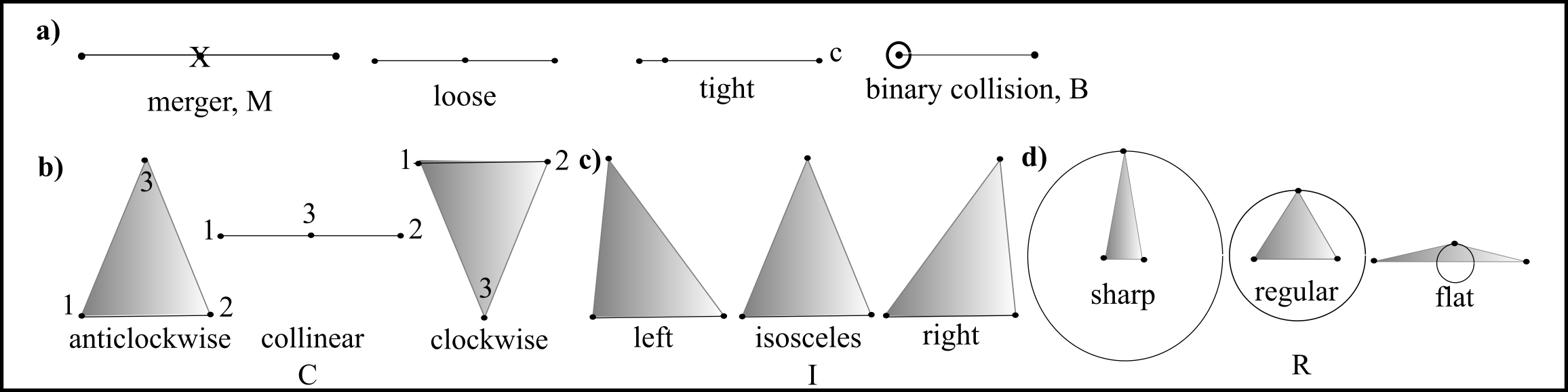}
\caption[Text der im Bilderverzeichnis auftaucht]{        \footnotesize{Types of configuration for 3 particles in a) 1- and b) 2-$d$. 
`Tight' is used here as in `tight binary' from Celestial Mechanics and Astronomy.} }
\label{Configs} \end{figure}          }
%
{            \begin{figure}[!ht]
\centering
\includegraphics[width=0.6\textwidth]{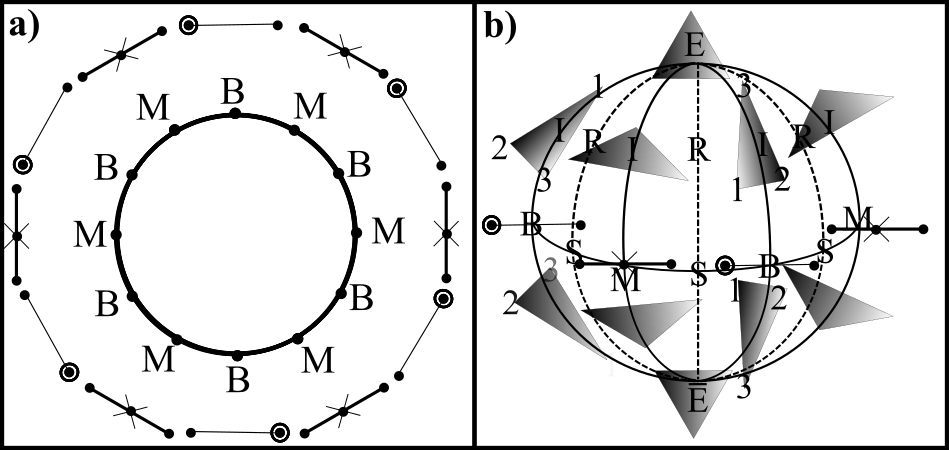}
\caption[Text der im Bilderverzeichnis auftaucht]{        \footnotesize{a) 3-stop metroland `clock face'. 

\m 

\n b) Triangleland `austroboreal zodiac' in 2-$d$ \cite{Kendall89, FileR, M15, S-III}. } }
\label{3-1-2} \end{figure}          }
%
\n{\bf Remark 1} See Fig \ref{Configs}.a) for different kinds of labelled mirror-image-distinct 3-stop-metroland configurations 
and Fig \ref{3-1-2}.a) for where these live on the shape circle; 
\cite{S-I} covers each of these matters in consummate detail. 

\m 

\n{\bf Remark 2} See Fig \ref{Configs}.b) for different kinds of labelled mirror-image-distinct triangles 
and Fig \ref{3-1-2}.b) for where these live on the shape sphere; 
now \cite{S-III} gives detailed coverage each of these topics.  

\m 

\n{\bf Remark 3} Such tessellations provide a useful `interpretational back-cloth' (c.f.\ Fig \ref{3-1-2}.b)) for the study of dynamical trajectories, 
                                                                                                                               probability distributions and 
																															   quantum wavefunctions.   
This method was originally applied in the Shape Statistics setting by Kendall \cite{Kendall89}: he termed this method `spherical blackboard'.  

\m 

\n{\bf Remark 4} $\mB$ denotes binary {\it collision} 
             and $\mM$ denotes        {\it merger}:    a configuration in which the third particle coincides with the centre of mass of the other two.
In 3-stop metroland, the 6 $\mB$ are evenly spaced at $\pi/3$ to each other, with the 6 $\mM$ at the midpoints of the arcs between these, 
thus forming overall the `clock face' exhibited in Fig \ref{3-1-2}.a).  
This imagery moreover succeeds in having 12 equally spaced distinctive points, 
i.e.\ the clock face's hour markers are labelled alternately with $\mB$ and $\mM$. 

\m 

\n{\bf Remark 5} The {\it equilateral triangles} $\mE$ and orientation-reversed $\overline{\mE}$ are at two antipodal points on the triangleland shape sphere, 
making for a very natural and significant choice of poles thereupon.  

\m 

\n{\bf Remark 6} The {\it collinear configurations} $\mC$ form the equator of collinearity on the triangleland shape sphere.
This divides the shape sphere into hemispheres of clockwise and anticlockwise oriented triangles, as per Fig \ref{Configs}.

\m

\n{\bf Remark 7} {\it Isosceles configurations} $\mI$ form 3 bimeridians at $2\pi/3$ to each other. 
{\it Regular configurations} $\mR$ are triangles with $I_{\sb\sa\sss\se} = I_{\sm\se\sd\si\sa\sn}$: equal base and median partial moments of inertia. 
These form the 3 bimeridians perpendicular to each of the three bimerdians of isoscelesness.  
That there are 3 types of isosceles $\mI$ and regular $\mR$ 
in correspondence to there being 3 ways of picking a base (and thus an apex and so a median from the base's centre to the apex).  
In contrast, equilaterality and collinearity are clearly labelling-independent notions. 
Bimeridians are indeed great circles (i.e.\ spherical geodesics) emanating moreover from the poles, 
so all 3 notions of regularity, as well as all 3 notions of isoscelesness are concurrently attained for equilaterality alone.  
    Each $\mI$ divides the shape sphere into hemispheres of left  and right leaning triangles, 
and each $\mR$ into                          hemispheres of sharp and flat          triangles, all as per Figs \ref{Configs}.b) and \ref{3-1-2}.b).

\m  

\n{\bf Remark 8} In triangleland, the $\mB$ lie at 120 degrees to each other on equator of collinearity, each with an antipodal $\mM$. 
Each such pair is where each bimeridian of isoscelesness $\mI$ intersects the equator of collinearity $\mC$.  
Thus along $\mC$, each $\mM$ bisects the arc between two $\mB$, and vice versa.   
The intersections of the bimeridians of regularity $\mR$ with the equator of collinearity $\mC$ have no salient further properties, however,  
so we denote these configurations by $\mS$ for `spurious'.  

\m 

\n{\bf Remark 9} The overall picture (Fig \ref{3-1-2}.b) is that of an orange of 12 segments cut into hemispheres perpendicular to those segments, 
as might well occur in preparing to `juice' the orange. 
While this depiction has no reason to favour 12 segments, the following further depiction does incorporate this.   
Namely, the triangleland shape sphere exhibits exactly the same pattern as the `austroboreal zodiac' of the celestial sphere. 
I.e.\ a partition into both the 12 signs of the zodiac and into Northern and Southern skies.
This partition being into equal pieces -- demilunes -- it constitutes moreover a {\it tessellation} alias {\it tiling} with these demilunes as tiles.  
Faces, edges and vertices therein being physically significant in the current context, one is really dealing with a {\sl labelled} tessellation.  

\m 

\n{\bf Remark 10} 4-stop metroland -- the relationally minimal unit for similarity shapes in 1-$d$ also has a spherical shape space admitting a tessellation. 
While there is nothing group-theoretically special about the (halved) 12-segment orange among its $p$-segmented kin, 
4-stop metroland itself realizes \cite{AF} a furtherly group-theoretically special \cite{Magnus} tesselation based on the cube--octahaedron group.  
\cite{S-II} gives a detailed study of 4-stop metroland. 
Finally, \cite{QuadI} is a first study of how quadrilateralland's $\mathbb{CP}^2$ is decorated, 
quadrilateralland being far more typical of an $N$-a-gonland than triangleland on account of its complex-projectiveness not reducing to sphericity.  

\m 
 
\n{\bf Remark 13} As mechanical theories, RPMs have positive-definite kinetic arc elements, which significantly differ from GR's indefinite one. 
[This Series of Articles other principal model arenas -- minisuperspace and SIC's perturbations thereabout -- do however inherit GR's indefiniteness.]

\subsection{GR's reduced configuration spaces}\label{RCS}
%
{            \begin{figure}[!ht]
\centering
\includegraphics[width=0.8\textwidth]{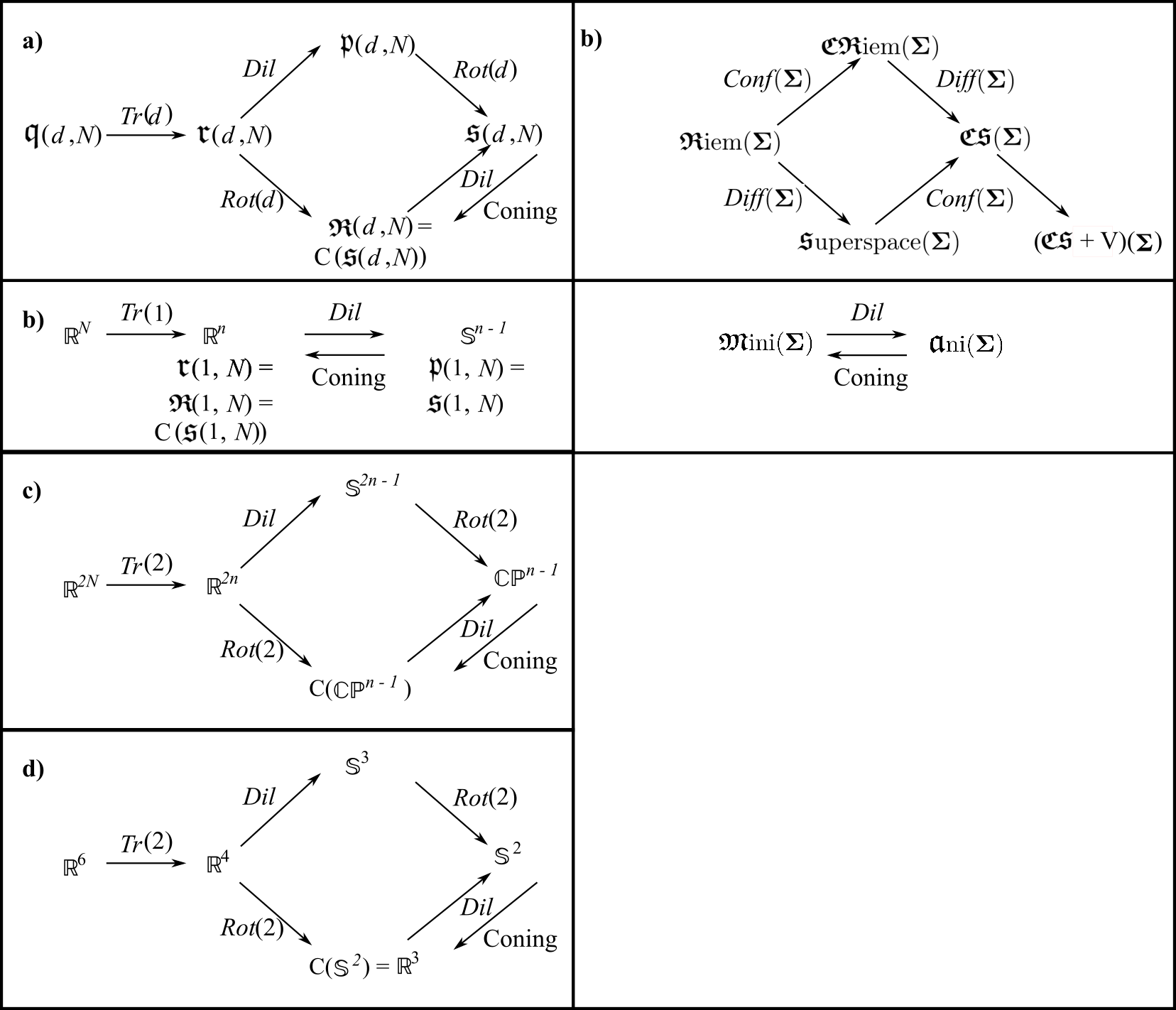}
\caption[Text der im Bilderverzeichnis auftaucht]{  \footnotesize{  This Section's configuration spaces a), 
                                                                    with b) 1-$d$, c) 2-$d$ and 3) triangleland specializations 
                                                                    are useful model arenas for their GR counterparts e) 
																	whose minisuperspace specialization is given in f).
																	See \cite{ABook, Minimal-N, A-Killing, A-CBI} for further examples. } }
\label{Q-RPM-GR} \end{figure}          }

\n GR's 3-geometry is a further notion of locally-scaled shape.  
Rotationally-invariant inner products are Euclidean RPM's analogues. 
Wheeler's superspace in this light is a locally-scaled shape space.

\m 

\n Fig \ref{Q-RPM-GR} also mentions Fischer and Moncrief's \cite{FM96} {\it conformal Riem} $\CRiem(\bupSigma)$ of GR as an analogue of Kendall's preshape space, 
as well as York's \cite{York74, FM96} more famous {\it conformal superspace} $\CS(\bupSigma)$ of GR as an analogue of Kendall's shape space. 
York's \cite{York72} $\CS + V$ just appends a single global degree of freedom -- the volume of the universe -- to this, 
the local versus global distinction of GR breaking Euclidean shapes-and-scales' return to relationalspace upon coning shape space.   
These further GR configuration spaces play a major role in its Initial-Value Problem \cite{York72}, out of permitting decoupled treatment of $\u{\sbcM}$ and $\scH$. 
They are associated with, respectively, maximal and constant-mean-curvature slices, which play a further role in Articles III and IX.

\section{Indirect implementation: the $\lFrg$-act $\lFrg$-all method}\label{Gact-Gall} 

\subsection{$\lFrg$-act $\lFrg$-all Method}  

In many cases, however, working directly is not possible since $\w{\FrQ}$'s geometry is unknown, 
                                                                                        complicated,
																				     or highly pathological (e.g.\ as per Article XIII).  
There is however a distinct indirect approach \cite{FileR, APoT3, ABook} which, at least formally, is {\sl universal}. 

\m 

\n{\bf Step 1} Given objects $\biO(\biQ)$ that are not $\lFrg$-invariant, firstly act on these with $\lFrg$, 
\be 
\s{\rightarrow}{\lFrg}_g \biO(\biQ)                                                               \m .
\ee
\n{\bf Step 2} Use some operation $\lS$ over all of $\lFrg$ to eliminate $\lFrg$-dependence: 
\be 
\biO_{\nFrg\mbox{-}\si\sn\sv}  \:=  \lS_{\sbig \m \in \m \nFrg} \s{\rightarrow}{\lFrg}_g  \biO(\biQ)  \m . 
\ee
This has the effect of cancelling out $\lFrg$-act's use of $\bigg$, thus producing overall a $\lFrg$-invariant version of $\biO$.

\m 

\n{\bf Naming 1} The true-name of this procedure is {\bf $\lFrg$-act $\lFrg$-all Method} \cite{FileR}. 

\m 

\n{\bf Definition 1}
\be
CR: \biO  \longrightarrow  \biO_{\nFrg\mbox{-}\si\sn\sv} 
%
%
\ee 
are {\it Configurational Relationalism implementing maps}.

\subsection{Scope}  

\n This method is universal over type of mathematical object $\biO$.

\m 

\n At the classical level, this includes composites solely of configurational variables, $\biO(\biQ \mbox{ alone})$, 
                                                                              as well as $\biO(\biQ, {\biQ}^{\prime} \mbox{ alone})$,  
																			             $\biO(\biQ, \dot{\biQ} \mbox{ alone})$,  
																						 $\biO(\biQ, \d{\biQ} \mbox{ alone})$ and 
																						  $\biO(\biQ, {\biP} \mbox{ alone})$.  

\m 

\n{\bf Example O1} Commonly-used such composites include Principles of Dynamics actions. 

\m 

\n{\bf Remark 1} It is important at this stage however to emphasize {\sl how much the scope has been broadened} away from previous Sections' such actions, 
with the following further examples covered by the $\lFrg$-Act $\lFrg$-All Method.

\m 

\n{\bf Example O2} Notions of distance \cite{Korner, Gromov}. 

\m 

\n{\bf Example O3} Notions of estimator \cite{CB, Kendall, PE16}.  

\m 

\n{\bf Example O4} Notions of information \cite{CT, ABook}.

\m 

\n{\bf Example O5} Notions of correlation \cite{H01, ABook}. 

\m 

\n{\bf Example O5} Quantum operators \cite{RS, ABook}.

\subsection{`All' operations. i) Measure of the Whole}  

\n One of the most elementary and familiar $\lFrg$-Act $\lFrg$-All Method uses {\it the measure of the whole}.
This means summing over a finite group 
\be 
\lS  \es  \sum_{\sbig \m \in \m  \nFrg}                    \m , 
\ee 
or integrating over a compact Lie group
\be 
\lS  \es  \int_{\sbig \m \in \m \nFrg} \d \bigg_{\sH\sa\sa\sr} \m \times \m \m , 
\ee
for Haar measure \cite{Kendall} $\d \bigg_{\sH\sa\sa\sr}$.

\subsection{ii) Group Averaging}

The other the most elementary and familiar examples of $\lFrg$-Act $\lFrg$-All Method is {\it group averaging}. 
This is named after its `all' operation the `all' operation, 
\be 
\lS \es \frac{1}{|\lFrg|}\sum_{\sbig \m \in \m  \nFrg} \m \mbox{ (finite groups)}                                                          \m , 
\ee 
or
\be 
\lS \es \frac{1}{\int_{\sbig \m \in \m \nFrg}\d \bigg_{\sH\sa\sa\sr}  } \, \int_{\sbig \m \in \m \nFrg} \d \bigg_{\sH\sa\sa\sr} \m \times \m \m  
                                                        \mbox{(compact Lie groups)}                                                      \m ,   
\ee
{\bf Remark 1} Cauchy \cite{Cauchy} is the first known proponent of group averaging, 
this being, moreover, the first use of any kind of $\lFrg$-act $\lFrg$-all Method.  
In particular, he used it in 1845 to prove a version of what is now known as Burnside's Lemma.
It was moreover with Burnside's own later work  \cite{Burnside} that group averaging entered common knowledge among mathematicians; 
see \cite{Serre, FHBook, Bala-Combi} for some basic modern uses. 

\m 

\n {\bf Remark 2} Group averaging is elementary enough to feature in undergraduate-level Group Theory and Representation Theory courses. 
It is thus very widely known among both Mathematics and Physics majors. 
We thus emphasize it as a more widely known and broadly applicable precursor of handling Configurational Relationalism 
than either the aforementioned Best Matching or its Thin Sandwich subcase.

\subsection{iii) Optimization}  

One can also optimize over the group $\lFrg$. 
In continuum cases, this can be approached by Calculus, whereas direct methods may be available regardless of continuum, discrete or intermediary status.
%

\m 

\n Under some circumstances, a global maximum or supremum may be required, 
\be 
\lS  \es  \stackrel{\mbox{sup}}{\sbig \in \nFrg} 
\ee 
or a global minimum or infimum.
\be 
\lS  \es  \stackrel{\mbox{inf}}{\sbig \in \nFrg} 
\ee 
In other continuum cases, a local maximum or minimum would do, or even some other type of critical point such as an inflexion, 
or an unqualified or unknown type of critical point.  
Then we use the notation 
\be 
\lS  \es  \lE_{\sbig \in \nFrg} 
     \:=  (\mbox{extremum of $\bigg$-dependent expression acted upon over $\lFrg$})
\ee 
The aforementioned Best Matching is a subcase of this.

\subsection{Inserting {\biM\bia\bip}s}

The $\lFrg$-Act $\lFrg$-All Method can moreover be further generalized 
by inserting any well-defined $\lFrg$-independent $Map$ between the $\lFrg$-act and $\lFrg$-all operations: 
\beq
CR(\biO) \:=  \biO_{\nFrg\mbox{-}\si\sn\sv} 
         \:=  \lS_{\sbig \, \in \,  \nFrg} \circ \, Map \, \circ \stackrel{\rightarrow}{\lFrg} \biO              \m . 
\eeq
Were $Map$ not $\lFrg$-invariant, we would take $\lFrg$ to act rather on a new type of object 
\be 
\biO^{\prime} = Map \circ \biO  \m . 
\ee 
\n`$Map$' moreover confers general usefulness in isolating the     $\lFrg$ part of an object under study in isolation 
                                                      from further $\lFrg$-invariant `paraphernalia factors' 
													  (c.f.\ isolation of parameter-dependent parts in the Principles of Statistics \cite{CB}).

\m 

\n{\bf Example 1} $Map$ is trivial in standard group averaging. 

\m 

\n{\bf Example 2} A $Map$ could moreover be inserted to extend this to {\it weighted} group averaging.  

\m 

\n{\bf Example 3} Best Matching can be thought of as involving 
\be
\times \sqrt{2 \, W} \m \mbox{ and } \m \int
\ee 
$Map$s acting on the second of Sec \ref{Core}'s core objects, schematically   
\beq
CR(\biO) \:=  \biO_{\nFrg\mbox{-}\si\sn\sv}  
         \:=  \lS_{\sbig \, \in \,  \nFrg} \circ \, Maps \, \circ \stackrel{\rightarrow}{\lFrg}_{\sbig} \biO  \m ,  
\eeq
now taking a specifically TRi-preserving form.

\m 

\n Let us end by giving some illustrative examples.

\subsection{Notions of distance on configuration spaces}\label{Dist}

\n We will use the following to construct various further examples of $\lFrg$-Act $\lFrg$-All.  

\m 

\n{\bf Structure 1} Various such \cite{Kendall84, Kendall, BB82, DeWitt70, FileR} can be built from 
the inner product and norm corresponding to the metric $\biM$  on configuration space $\FrQ$:   
\be
\mbox{(Kendall Dist)} = (\biQ, \biQ^{\prime})\mbox{}_{\sbiM} \m ,    
\label{ProtoKenComp}
\ee
which compares two finitely separated configurations. 
\be
\mbox{(Barbour Dist)} = ||\d \biQ||_{\sbiM}\mbox{}^2 \m , 
\label{ProtoBarComp}
\ee
which considers the difference between infinitesimally separated configurations, $\biQ$ and $\biQ + \d \biQ$.  
\be
\mbox{(DeWitt Dist)} = (\d \biQ, \d \biQ^{\prime})_{\sbiM} \m ,   
\label{ProtoDeWittComp}
\ee
which compares two differences between infinitesimally separated configurations. 

\m 

\n{\bf Remark 1} For all that we use a norm symbol above, this is only a bona fide norm of $\biM$ is positive-definite,
which applies to Shape(-and-Scale) Theory but neither to GR nor to requisitely $\geq 2$ degrees of freedom minisuperspace models.  
In these other cases, the separation property of distance is lost, for the well-known reason that null directions are supported.

\subsection{Kendall, Barbour and DeWitt comparer cores}\label{Core}

\n{\bf Structure 1} Next suppose that there is an $\lFrg$ acting upon $\FrQ$ that is held to be physically irrelevant 
(or, more generally, irrelevant to the modelling in question). 
In this way, we have the following $\lFrg$-act cores:    
\be
\mbox{(Kendall $\lFrg$-Dist)} = (\biQ \cdot \stackrel{\longrightarrow}{\lFrg_{g}} \biQ^{\prime})_{\sbiM} \m ,  
\label{Kend}
\ee 
\be
\mbox{(Barbour $\lFrg$-Dist) } =  ||\d_{\sg}\biQ||_{\sbiM}\mbox{}^2 \m \mbox{ and }  
\label{Barb}
\ee
\be
\mbox{(DeWitt $\lFrg$-Dist)}   =  (\stackrel{\longrightarrow}{\lFrg}_{\d g} \biQ, \stackrel{\longrightarrow}{\lFrg}_{\d g} \biQ^{\prime})_{\sbiM}   \m .		
\label{DeWi}
\ee
{\bf Remark 1} The first and third having two arguments, one might contemplate acting on the other argument instead. 
But our objects are symmetric, so this makes no difference. 

\m 

\n Acting on both arguments is unnecessary if {\it equivariance} applies \cite{Younes}; for a map $\phi$, this means that 
\be
\phi(g \, x) = g \, \phi \, x   \m \m \forall
 g \in \lFrg \mma \forall x \in X \m \mbox{ (the acted-upon set)}  \m .  
\ee

\subsection{Kendall's comparer}

{\bf Structure 1} This involves pairing Kendall's $\lFrg$-act core with an extremizing `all' operation. 
Kendall thus obtained the following comparer between similarity shapes \cite{Kendall84, Kendall89, Small, Kendall, Bhatta, PE16}:  
\be
\mbox{(Kendall $\lFrg$-Dist)}  \es  \stackrel{\mbox{min}}{\mbox{\scriptsize $\u{\theta} \m \in \m SO(d)$}} \m \u{n} \, \u{\u{Rot}}_{\u{\theta}} \, \u{n}    \m   
\label{Kend-3}
\ee
for normalized centre-of-mass frame preshape vectors $\u{n}$.  

\m 

\n{\bf Remark 1} Kendall used the above to attain minimal incongruence between planar figures. 
Contrast this description with Best Matching's, for all that Kendall's notion involves just configurations 
rather than any notion of change as enters Best Matching itself.

\subsection{DeWitt's comparer}

{\bf Structure 1} The arena in which DeWitt proposed his comparer is GR-as-Geometrodynamics. 
He considered the double-multiple integral, over primed and unprimed coordinates associated with the first and second 3-metric being compared.  
This bears various similarities to the BSW action, but is not equivalent to it, as explained in Sec \ref{Comparison}.

\subsection{Barbour's Best Matching comparer}

\n As alluded to before, this involves firstly taking Barbour's core object and applying $\times \sqrt{2 \, W}$ and integration $Maps$ 
to form a Principles of Dynamics action ${\cal S}_{\sJ}$, before finally partnering this with a $\lFrg$-all extremum move.  
The bare object here is of $\biO(\biQ, \dot{\biQ})$ or $\biO(\biQ, \d \biQ)$ type, depending on implementation as per Article I's distinction. 

\m  

\n{\bf Remark 1} This can furthermore now be recognized also as a subcase of {\it weighted path metric}: a standard Analysis construct \cite{Korner} 
(modulo separation, in the indefinite case).     
It is the $\sqrt{2 \, W}$ factor that plays the role of the weighting function.  

\m

\n{\bf Remark 2} Best Matching versus $\lFrg$-Act $\lFrg$-All gives our Series second distinction between Small and Large Methods respectively.

\subsection{Our comparers contrasted}\label{Comparison}

\n{\bf Remark 1} (\ref{Kend}) itself differs from the other two cases in using a {\sl finite} group action to the other two cases' infinitesimal ones.  
Indeed, the geometrodynamical analogue of Kendall's construct can be envisaged to give the core of the Thick Sandwich.  

\m 

\n{\bf Remark 2}  In another sense, it is (\ref{Kend}) and (\ref{DeWi}) which are akin: 
these compare two distinct inputs whereas (\ref{Barb}) works with a single input.  
This corresponds to Barbour's comparer being `thin', in the Thin Sandwich sense in the case of GR-as-Geometrodynamics, 
albeit extending also to any other theory's Best Matching being a `thin' construct.

\m 

\n{\bf Remark 3} (\ref{DeWi}) and (\ref{Barb}) differ in the manner of which order roots and sums-or-integrals are taken;  
this is a prototype of how Jacobi--Synge actions generalizes Jacobi ones (Sec I.4.6).  

\m 

\n{\bf Remark 4}  If $\biM = \biM(\biQ)$, does one use the first configuration $Q_1$ or the second $Q_2$ in evaluating $\biM$ itself? 
This situation does not arise for Kendall's shapes in $\mathbb{R}^n$, but it does in DeWitt's geometrodynamical context. 
DeWitt resolved this in the symmetric manner: using $\bih$ and $\bih^{\prime}$ to equal extents; 
this is possible by the GR configuration space metric being quadratic in the spatial metric.

\subsection{The Fischer Comparer}

$\Superspace(\bupSigma)$ additionally admits a metric space metric of the form \cite{Fischer70} 
\beq
\mbox{Dist}([\bh_1], [\bh_2]) := \inf_{\phi \, \in \, \mbox{\scriptsize $Diff$}(\sbupSigma)}\left(\mbox{Dist}(\phi^*\bh_1, \phi^*\bh_2)\right) \mbox{ } .
\label{Fischer-Dist}
\eeq
I.e.\ a $Diff(\bupSigma)$ specific realization of Kendall's `act' core paired with the inf `all' move.

\subsection{Younes Comparers}

These \cite{Younes} are also based on a Kendall `act' core with diffeomorphisms acting thereupon. 
The objective now is to match pieces of curves to each other, or of surfaces to each other, i.e.\ local submanifold matching within a given geometry.
This is attained with a variational extremization `all' move. 
Vector field and frame field matching versions of this are also available.

\subsection{The Hausdorff Comparer}

{\bf Structure 1} The core object here is the {\it Hausdorff distance} \cite{Gromov}  
\be
\mD_{\sH}(X, Y) = \mbox{sup}(\d(x, Y) | x \in X	)
\ee 
$\d$ is here a metric, our aim being to evaluate distance between subsets of a given metric space. 

\m 

\n{\bf Structure 2} Our `Act' move is then on one one argument of this, followed by the inf `All' move to create the {\it Hausdorff comparer}
\be 
\stackrel{\mbox{inf}}{\mbox{\scriptsize{$I \in Isom(Y)$}}} \m D_{\sH}(X, I(Y))  \es  
\stackrel{\mbox{inf}}{\mbox{\scriptsize{$I \in Isom(Y)$}}} \m \mbox{sup} (\d(x, I(Y)) | x \in X	)  \m .  
\ee 	
\n{\bf Remark 1} A significant and furtherly modern application of this is that it enters the definition of 
{\it Gromov--Hausdorff distance} of deviation from isometry between compact metric spaces.

\subsection{Contrast with direct and piecewise-intrinsic comparers}

\n{\bf Remark 1} In some cases, one might instead be able to work directly with, or reduce down to, $\w{\FrQ}$ objects, 
in which case there is no need for the above indirect construct. 
One would then make use of the relational or reduced $\w{\FrQ}$ geometry $\w{\biM}$ itself.

\m 

\n{\bf Remark 2} An alternative to the above comparisons of two configurations themselves is to performing intrinsic computations on each 
\beq
\mbox{\Large $\iota$}: \FrQ \longrightarrow \mathbb{R}^{p} \m ,                                                                                        
\eeq
and only then compare the outputs of these computations.\footnote{This is motivated e.g.\ by some of the preceding comparers failing to give distances when $\mbox{\boldmath$M$}$ 
is indefinite -- losing the non-negativity and separation properties of bona fide distance -- which we know will occur for GR.}

\m 

\n{\bf Remark 3} One can now consider using norms in the space of computed objects that it is mapped into (the $p$-dimensional Euclidean metric in the above example).  
The outcome of doing this may however depend on the precise quantity under computation. 

\m 

\n{\bf Remark 4} $\mbox{\Large $\iota$}$ will in general has a nontrivial kernel. 
So the candidate $\mbox{\Large $\iota$}$-Dist would miss out on the separation property of bona fide distances.  
If this separation fails, one can usually (see e.g.\ \cite{Gromov}) quotient so as to pass to a notion of distance. 
[Occasionally this leaves one with a single object so that the candidate notion of distance has collapsed to a trivial one.]

\m 

\n{\bf Remark 5} It is however limited or inappropriate to use such a distance if it is the originally intended space 
rather than the quotient that has deeper significance attached to it.  

\m

\n Examples of $\mbox{\Large $\iota$}$ include inhomogeneity quantifiers, 
                                               spectral quantities \cite{Berger-rama}, 
                                      and some instances of `best metrics' \cite{Berger-rama}.  

\m 

\n{\bf Remark 6} $\mbox{\Large $\iota$}$ can again be directly or indirectly $\lFrg$-invariant.

\m 

\n{\bf Selection Criterion 1} Indeed, amongst the vast swathe of possibilities for $\mbox{\Large $\iota$}$, 
or Configurationally Relational constructs more generally, {\sl directness} is itself a selection criterion.

\m 

\n{\bf Selection criterion 2)} `{\sl Physical naturality}', i.e.\ recurrence of the structure used in other physical computations.
E.g.\ a notion of distance that is -- or at least shares structural features with -- 
an action, Statistical Mechanics partition function, entropy, or notion of information,  or a quantum path integral.  
This criterion's scope can moreover be extended to `{\sl modelling naturality}' in other contexts such as Statistics or Geometry. 

\m 

\n{\bf Selection criterion 3)} {\sl Extendibility to unions of configuration spaces}, 
so to permit inclusion of `topology change', `big Superspace' \cite{Fischer70}, or other arenas' analogues (Epilogues II.B-C of \cite{ABook}). 

\m 

\n{\bf Remark 7} This is not an exhaustive list of selection criteria, with 2) and 3) being intended to serve as selection principles for 
specifically $\lFrg$-Act $\lFrg$-all constructs implementing Configurational Relationalism. 

\m 

\n{\bf Remark 8} Notions of information and of correlation can be developed in comparable detail to how distance is treated above; 
see Appendices Q and U of \cite{ABook} for an outline.

\section{Conclusion}
%
{            \begin{figure}[!ht]
\centering
\includegraphics[width=1\textwidth]{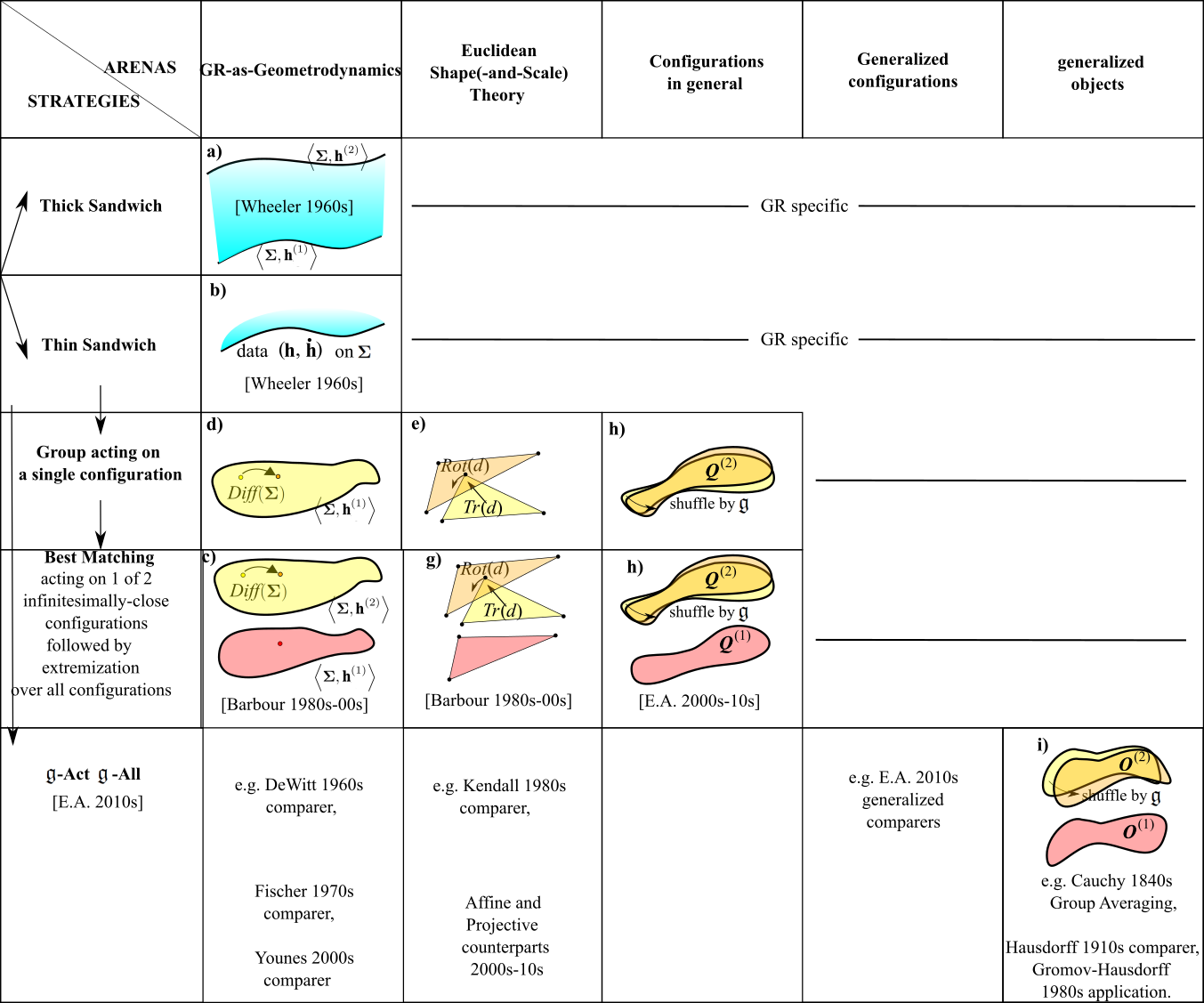}
\caption[Text der im Bilderverzeichnis auftaucht]{        \footnotesize{End-summary of progression of strategies for Configurational Relationalism.

\m 

\n c) The Thin Sandwich can now be reinterpreted in terms of Best Matching $\Riem(\bupSigma)$ with respect to $Diff(\bupSigma)$.
In c's depicted shuffling, the red space is held fixed while each point in the yellow space is moved to a new position marked in orange 
so as to seek out minimal incongruence between the two.  
%

\m 

\n This rests on d)'s more basic point that $Diff(\sbupSigma)$ acts on $\sbupSigma$ by moving points around.   } }
\label{CR-Development} \end{figure}          }

\n For now, we provide a summary figure \ref{CR-Development} for the development of Configurational Relationalism, 
leaving it to Article IV to jointly conclude for the piecemeal-facet Articles I to IV.   


\end{document}